\documentclass[AMA,STIX1COL]{WileyNJD-v2}

\articletype{Research Article}%

\received{26 April 2016}
\revised{6 June 2016}
\accepted{6 June 2016}

\begin{document}

\title{An Integral Sliding-Mode Parallel Control Approach for General Nonlinear Systems via Piecewise Affine Linear Models}

\author[1,2]{Chunyang Zhang}

\author[1,2]{Qing Gao*}

\author[3]{Yue Deng}

\author[4,5]{Jianbin Qiu}

\authormark{Chunyang Zhang \textsc{et al}}

\address[1]{\orgdiv{School of Automation Science
		and Electrical Engineering}, \orgname{Beihang University}, \orgaddress{\state{Beijing}, \country{P.R. China}}}

\address[2]{\orgname{Zhongguancun Laboratory},  \orgaddress{\state{Beijing}, \country{P.R. China}}}

\address[3]{\orgdiv{School of Astraunotics}, \orgname{Beihang University}, \orgaddress{\state{Beijing}, \country{P.R. China}}}

\address[4]{\orgdiv{ State Key Laboratory of Robotics and Systems}, \orgname{Harbin Institute of Technology}, \orgaddress{\state{Harbin}, \country{P.R. China}}}

\address[5]{\orgdiv{Research Institute of Intelligent Control and Systems}, \orgname{Harbin Institute of Technology}, \orgaddress{\state{Harbin}, \country{P.R. China}}}

\corres{*Qing Gao, School of Automation Science
	and Electrical Engineering, Beihang University, Beijing 100191, P.R. China. \email{gaoqing@buaa.edu.cn}}


\abstract[Summary]{The fundamental problem of stabilizing a general non-affine continuous-time nonlinear system is investigated via piecewise affine linear models~(PALMs) in this paper. A novel integral sliding-mode parallel control~(ISMPC) approach is developed, where an uncertain piecewise affine system~(PWA) is constructed to model a non-affine continuous-time nonlinear system equivalently on a compact region containing the origin. A piecewise sliding-mode parallel controller is designed to globally stabilize the PALM and, consequently, to semi-globally stabilize the original nonlinear system. The proposed scheme enjoys three favorable features: i) some restrictions on the system input channel are eliminated, thus the developed method is more relaxed compared with the published approaches;  ii) it is convenient to be used to deal with both matched and unmatched uncertainties of the system; and (iii) the proposed piecewise parallel controller generates smooth control signals even around the boundaries between different subspaces, which makes the developed control strategy more implementable and reliable. Moreover, we provide discussions about the universality analysis of the developed control strategy for two kinds of typical nonlinear systems. Simulation results from two numerical examples further demonstrate the performance of the developed control approach.}

%

\maketitle

\footnotetext{\textbf{Abbreviations:} PALMs, piecewise affine linear models; ISMPC, integral sliding-mode parallel control; PWA, piecewise affine system; LMIs, linear matrix inequalities; GAS/GES, globally asymptotically/exponentially stabilizable.}

\section{Introduction}
\label{sec:tro}
PALMs possess the convenience of control design and  simplicity of structure, which have resulted in its extensive employment in analysis and control of diverse industrial systems with nonlinearity~\cite{johansson2003piecewise,kersting2017direct,li2018stability}. By dividing the premise state space into a series of adjacent subspaces, PALMs model a nonlinear system equivalently by an affine linear system with norm-bounded uncertainties that can be made small enough via  appropriate design in each subspace~\cite{julian1998canonical}. Based on the powerful linear system theory and a quadratic Lyapunov function, this comparatively simple framework promotes systematic analysis and controller synthesis for nonlinear systems significantly and  fruitful research in this field have been published in the past
decades~\cite{qiu2011approaches,zhang2016observer,wei2018new,desimini2020robust,xu2022passivity}.

As another research frontier in the robust control theory, sliding-mode control (SMC)~\cite{utkin1977variable} has drawn growing research interests and has been used in many industrial applications~\cite{razmi2019neural,wang2019new,wang2020usde}. The SMC strategy holds various favorable characteristics, like unique robustness against disturbances and uncertainties, distinguished transient performance and fast response. The core idea
of the SMC approach is to construct the closed-loop control
system such that the system trajectories are first driven onto a well-designed linear sliding surface covering the equilibrium, and are forced to maintain on the surface with preferred
convergent characteristic towards the equilibrium~\cite{LIU2020108596,wei2021dynamic,zhang2021design,zhang2021finite}. The special dynamics the closed-loop control system
behaves while its trajectories are moving on the sliding surface
is called the sliding motion. An alternative approach is the integral sliding-mode control (ISMC) scheme where an integral form sliding surface is used instead of a linear one. Different from the common SMC scheme, in the
ISMC approach, the system trajectories maintain on the sliding
surface during the whole time interval~\cite{utkin1997integral}. Consequently,
the reaching phase can be removed from the dynamics w.r.t. the
controlled system, which demonstrates stronger robustness of
the ISMC approach than the SMC approach~\cite{gao2014suniversal,pan2017integral,gao2019fault,li2021a,ZHANG2022491}. 

By now, there have been few published results on ISMC design for PALMs. Considering the equivalence between a T-S fuzzy system and an uncertain PALM~\cite{cao1997analysis}, a convenient extension might be made such that the fuzzy ISMC strategy in~\cite{ho2007robust} can be applied to PALMs. However, the fuzzy ISMC strategy in~\cite{ho2007robust} suffers from a restrictive assumption that each submodel of the fuzzy system must hold an identical constant input matrix, thus it is confronted with significant conservativeness when being applied to general nonlinear systems. Various attempts have been made to weaken or remove this restrictive assumption~\cite{rubagotti2011integral,li2013adaptive,jiang2018novel}. The approach in~\cite{rubagotti2011integral} replaced this assumption by a less conservative assumption in SMC design for fuzzy systems. A piecewise ISMC approach allowing different local input matrices can be found in~\cite{xi2010piecewise} where the region of interest was split into a series of subspaces and unique integral sliding surface was constructed w.r.t. all subspaces. However, the high complexity of the approach in~\cite{xi2010piecewise} obstructs its wide implementation in practice. 

Compared with the T-S fuzzy model based approach in~\cite{gao2013universal}, a PALM models a general nonlinear system using linearization method~\cite{teixeira1998stabilizing}, which yields fewer plant rules in general and thus less conservative controller design. Nevertheless, the affine terms appearing in the local models of the PALM, on the other hand, lead to more complicated analysis and synthesis. Moreover, the commonly used piecewise static feedback controllers for PALMs~\cite{rodrigues2005piecewise} suffer from abrupt changes around the boundaries between different subspaces due to the switching behaviors of the systems. How to avoid this undesired chattering phenomenon needs further investigation.

In this paper, motivated by the previous fuzzy-model-based result in~\cite{gao2013universal}, an appropriate ISMC design for general non-affine continuous-time nonlinear systems through PALMs is developed to eliminate the restriction on system matrices and to avoid the chatter phenomenon around boundaries between subregions. Specifically, a PALM is constructed on a compact region to express a controlled nonlinear system first and then an ISMPC method is proposed for global asymptotic stabilization of the PALM and, consequently, for ensuring the  semi-global asymptotic stability of the original general nonlinear system. In particular, the constructed integral type sliding surface function depends on the system state and control signal. Compared with the commonly designed static state feedback controller~\cite{rubagotti2011integral,li2013adaptive,jiang2018novel}, the ISMPC strategy utilizes a new parallel control law and, in each partition of the whole
system space, the resultant sliding-mode controller has a dynamical parallel
compensator form. The corresponding control gains
are obtained by calculating a series of linear matrix inequalities~(LMIs) with the aid of a common quadratic Lyapunov function. This ISMPC strategy holds three favorable features:

(i) Different input matrices w.r.t. the PALMs are allowed, thus the ISMPC scheme is applicable for general non-affine nonlinear systems;

(ii) The uncertainties arising during the approximation procedure, either ``matched'' or ``unmatched'', are eliminated in the control channel w.r.t. the resultant controlled system, which introduces stronger robustness;  and 

(iii) The proposed piecewise sliding-mode parallel controller admits a time-integral form of  solution and naturally generates smooth control signals, even around the boundaries of the partitioned subspaces where the controller gains switch abruptly. This helps reduce chattering phenomenon and makes the control law more implementable and reliable in real applications.

The universality discussion w.r.t. the developed ISMPC scheme in this paper is another key contribution. The key concern is that, for any given stabilizable general nonlinear system with
a smooth system function, can one always construct a piecewise integral
sliding surface and a corresponding piecewise integral sliding-mode
parallel controller such that the resultant closed-loop
control system behaves a stable sliding motion since initially? This universality characteristic of the proposed ISMPC strategy is analyzed for two classes of typical continuous-time nonlinear systems, i.e., globally asymptotically/exponentially stabilizable~(GAS/GES) nonlinear systems, respectively. It is believed that these discussions may provide confidence in applying the developed approach to wider industrial practice. 

The rest of this paper is structured as:
Section~\ref{sec:pre} formulates the PLAM and problems.
In Section~\ref{sec:ISMPC}, an ISMPC strategy is developed to globally
robustly stabilize a PALM and, correspondingly, to
semi-globally stabilize the corresponding original nonlinear
system on the predefined system space, then, the universality
discussion of the developed ISMPC scheme for~GAS/GES non-affine nonlinear
systems are presented respectively. The numerical simulation is implemented in Section~\ref{sec:sims}. Conclusion lies in Section~\ref{sec:conclusion}.

Notations: 
The notation $\star$ in a matrix expresses the entries induced by matrix symmetry.  Given a vector or a matrix $Q$, then $Q^T$ and $\lVert Q \rVert$ denote its transpose and induced norm respectively, and $Q>0$ indicates a matrix $Q$ is positive definite. Let $\alpha :\left[ 0,a \right) \rightarrow \left[ 0,\infty \right) $ be a continuous function, then $\alpha$ belongs to
the class $\mathcal{K}$, if $\alpha\left( 0 \right) =0$ and it is strictly increasing; $\alpha$ belongs to the class $\mathcal{K} _\infty$, if $a\rightarrow \infty$ as $t \rightarrow \infty$ and it belongs to the class $\mathcal{K}$.

%

\section{System Description and Preliminaries}
\label{sec:pre}
This paper will principally concentrate on the following general non-affine continuous-time nonlinear system defined on a
compact region $X \times U \subset \Re^{n} \times \Re^{m}$:
\begin{equation}
	\label{nonl}
	\dot{x}\left( t \right) =f\left( x\left( t \right) ,u\left( t \right) \right)
\end{equation}
where $x:=\left[ x_1 ,...,x_n \right] ^T \in X $, $u :=\left[ u_1 ,...,u_m \right] ^T \in U $. The following assumption is necessary: 

\textbf{Assumption 1.}

1). $X\times U$ contains the origin.

2). The system function $f$ is continuously differentiable on
$X\times U$ and has the origin as its equilibrium.
\label{assump2.1}

In view of the universal approximation capability of PALMs~\cite{julian1998canonical}, the following PALM can be constructed to express the controlled system in~(\ref{nonl}) equivalently on $X \times U $:
\begin{gather}
	\emph{for}\,\, \bar{x}\left( t \right) \subseteq \bar X_i=X_i\times U_i, i\in \varphi :=\left\{ 0,1,2,...,l \right\} \notag \\
	\dot{x}\left( t \right) =\left( A_i+\varDelta A_i(t) \right) x\left( t \right) +\left( B_i+\varDelta B_i(t) \right) u\left( t \right) +C_i+\varDelta C_i(t)
	\label{pls}
\end{gather}
where
\begin{gather}
	\label{erro1}
C_0=\varDelta{C}_0(t)\equiv 0, \lVert \varDelta C_j \rVert \leqslant \varepsilon _g,\lVert \left[ \varDelta A_0(t), \varDelta B_0 (t)\right] \rVert \leqslant \varepsilon _{f_0},\lVert \left[ \varDelta A_j(t),\varDelta B_j (t)\right] \rVert \leqslant \varepsilon _f, j\in \varPhi :=\left\{ 1,2,...,l \right\}, 
\end{gather}
$\bar{x}\left( t \right) =\left[ x^T\left( t \right) ,u^T\left( t \right) \right] ^T$, the norm-bounds of the approximation error $\varepsilon _{f_0}$, $\varepsilon _f$, and $\varepsilon _g$ can be made arbitrarily small~\cite{julian1998canonical}, $X_i$ and $U_i$ are the partitions of $X$ and $U$, respectively, $\bar X_i=X_i \times U_i$ are the adjacent partitions of $X\times U$, $l+1$ denotes the number of partitions, and $\bar X_0$ contains the origin.

Note that the partitions w.r.t. the PALM in~(\ref{pls}) are inherently polyhedral regions on the compact region.  To outer approximate the polyhedral regions $\mathcal{P}_{i}$, an ellipsoid $\mathcal{F}_{i}$ is used. Assume that the matrices $Q_i$ and $f_i$ can be designed to satisfy
\begin{align}
	\label{par1}
	\mathcal{P}_i\subseteq \mathcal{F}_i,\mathcal{F}_i=\left\{ \bar{x}|\lVert Q_i\bar{x}+f_i \rVert \leqslant 1 \right\} .
\end{align}

Suppose the polyhedral regions $\mathcal{P}_{i}$ are slabs, which are appropriately described as
\begin{align}
	\label{par2}
	\mathcal{P}_i=\left\{\bar{x}|\sigma  _{i_1}\leqslant \theta _{i}^{T}\bar{x}\leqslant \sigma  _{i_2} \right\},
\end{align}
where  $\sigma  _{i_1}\in \Re$, $\sigma  _{i_2} \in \Re$, and $\theta _{i} \in \Re^{n \times 1}$. Then each region can be precisely illustrated as a degenerate ellipsoid in~(\ref{par1}) with
\begin{align}
	\label{par3}
	Q_i=\frac{2\theta _{i}^{T}}{\sigma  _{i_2}-\sigma  _{i_1}},
	f_i=-\frac{\sigma  _{i_2}+\sigma  _{i_1}}{\sigma  _{i_2}-\sigma  _{i_1}}.
\end{align}

Based on~(\ref{par1}), the state $\bar{x}\left( t \right)$ within an ellipsoid region $\mathcal{F}_i$ satisfies
\begin{align}
	\label{par4}
	\left[ \begin{array}{c}
		\bar{x}\left( t \right)\\
		1\\
	\end{array} \right] ^{T}\left[ \begin{matrix}
		Q_{i}^{T}Q_i&		Q_{i}^{T}f_i\\
		\star&		f_{i}^{\text{T}}f_i-1\\
	\end{matrix} \right] \left[ \begin{array}{c}
		\bar{x}\left( t \right)\\
		1\\
	\end{array} \right] \leqslant 0.
\end{align}

It is noted that the PALM in~(\ref{pls}) approximates a smooth nonlinear system via linearization at multi-operating points in both the system state space and the control space~\cite{teixeira1998stabilizing}. However, some prior knowledge of the system behavior, which is often very difficult to obtain for complicated systems, is essential in this modeling process. One can refer to~\cite{cao1997analysis} for an approach to identifying these operating points via clustering algorithms. With the linearization points determined, the system space can be partitioned into a series of slab subspaces in~(\ref{par2}), each one of which envelops an operating point~\cite{johansson2003piecewise}, and can be formulated as in~(\ref{par4}).

By regarding the approximation errors arising during the modeling procedure as norm-bounded uncertainty terms, one can conclude that an ISMC scheme robustly stabilizing the PALM in~(\ref{pls}) can stabilize the original nonlinear system in~(\ref{nonl}) simultaneously.  Considering the equivalence between a T-S fuzzy system and an uncertain PALM~\cite{cao1997analysis}, one may extend the fuzzy ISMC approach in~\cite{rubagotti2011integral,li2013adaptive,jiang2018novel} to our case. However, the approaches in~\cite{rubagotti2011integral,li2013adaptive,jiang2018novel} are useful only when the nonlinear system in~(\ref{nonl}) has a constant and linear input channel. This motivates us to develop a new ISMC scheme that stabilizes the general non-affine nonlinear system in~(\ref{nonl}) based on its corresponding PALM in (\ref{pls}) and to remove these restrictions.

\section{Integral Sliding-Mode Parallel Control}
\label{sec:ISMPC}
A new ISMPC strategy will be presented in this section to robustly stabilize the PALM in~(\ref{pls}) and,  correspondingly, to stabilize 
the nonlinear system in~(\ref{nonl}).

\subsection{An Integral Sliding-Mode Parallel Controller}
\label{sec:DISS}
Considering the PALM in~(\ref{pls}), or equivalently~(\ref{nonl}), we propose a novel piecewise integral sliding surface as
\begin{align}
	&\quad\quad\quad\qquad \quad\textit{for}\,\, \bar{x}\left( t \right) \subseteq \bar X_i, i\in \varphi \notag \\
	s\left( t \right) =&S_x\left[ x\left( t \right) -x\left( 0 \right) \right]-\int_0^t{S_x\left( A_ix\left( s \right) +B_iu\left( s \right) +C_i \right)}ds \notag \\
	+&S_u\left[ u\left( t \right) -u\left( 0 \right) \right]-\int_0^t{S_u\left( F_ix\left( s \right) +G_iu\left( s \right) +D_i \right)}ds,
	\label{surf}
\end{align}
where $S_x\in \Re ^{m\times n}$ and $S_u\in \Re ^{m\times m}$ represent the sliding surface matrices to be designed and $S_u$ is required to be nonsingular. The matrices $F_i\in \Re ^{m\times n}$, $G_i\in \Re ^{m\times m}$ and $D_i\in \Re ^{m\times 1}$ will be determined later and here we set $D_0 \equiv 0$. 

The following theorem provides an appropriate design of the sliding-mode control law to guarantee that the integral sliding
surface in~(\ref{surf}) can be maintained from the beginning of evolution.

\begin{theorem}
	For the PALM in (\ref{pls}), or correspondingly,
	the controlled nonlinear system in (\ref{nonl}), by insulting a piecewise
	 sliding-mode parallel controller as
\begin{gather}
	\textit{for}\,\, \bar{x}\left( t \right) \subseteq \bar X_i, i\in \varphi \notag \\
	\dot{u}\left( t \right) =F_ix\left( t \right) +G_iu\left( t \right) +D_i -\left(\gamma + \alpha  _i+\upsilon _i\left( t \right) \right) S_{u}^{-1}\mbox{sgn}\left( s\left( t \right) \right)
	\label{cont}
\end{gather}
	with $u \left(0\right) = 0$,
\begin{equation}
	\label{errd}
	\alpha  _0=0, \alpha  _j=\varepsilon _g\lVert S_x \rVert,
	\upsilon   _0\left( t \right) =\varepsilon _{f_0}\lVert S_x \rVert \lVert \left[ x^T\left( t \right) ,u^T\left( t \right) \right] ^T \rVert ,
	\upsilon  _j\left( t \right) =\varepsilon _f\lVert S_x \rVert \lVert \left[ x^T\left( t \right) ,u^T\left( t \right) \right] ^T \rVert , j\in \varPhi
\end{equation}
	where $F_i$ and $G_i$ are defined in~(\ref{surf}), $\gamma>0$ is a scalar, the norm-bounds of the approximation error $\varepsilon _{f_0}$, $\varepsilon _f$ and $\varepsilon _g$ denote in (\ref{erro1}), then the piecewise integral sliding surface in (\ref{surf}) is
	reached and maintained since initially \emph{in potential}.
	\label{thm1}
\end{theorem} 

\begin{remark}
	Notice that the equivalence between the PALM in~(\ref{pls}) and the  continuous-time nonlinear system in~(\ref{nonl}) is ensured only within $X\times U$. Put another way, the  closed-loop control system consisting of (\ref{nonl}) and (\ref{cont}), which is named the \emph{practical closed-loop control system} in this paper, behaves the sliding motion since initially only when its trajectories keep moving within $X\times U$ during the time interval of interest. This is, however, often not the case in practice even for a stable closed-loop control system. When the initial states are very close to the boundary of $X\times U$, the system trajectories are highly possible to move out $X\times U$. In this case, the PALM in (\ref{pls}) and the piecewise integral sliding surface in (\ref{surf}) are both undefined, thus the developed ISMPC approach no longer works. Therefore, it is stated that by using the proposed approach, the ideal sliding mode can be only realized ``in potential". This will be illustrated by Fig. \ref{ex:illustrate} in the simulation section. Practically, designing a compact region where the PLAM is constructed big enough can improve this situation. 
	\label{remark3.1}
\end{remark}

\begin{remark}
The designed sliding-mode parallel control law in~(\ref{cont}) is in form of a dynamical parallel compensator, which distinguishes the proposed scheme from the published feasible solutions in~\cite{rubagotti2011integral,li2013adaptive,jiang2018novel}. One can observe an important advantage of this ISMPC strategy is that it can be applied to the PALM in~(\ref{pls}) without requiring that each local model holds an identical input channel, while in this general case, the methods in~\cite{rubagotti2011integral,li2013adaptive,jiang2018novel} cannot be directly used.
\end{remark}

\textbf{\emph{Proof}}:~The Lyapunov function candidate of
the piecewise integral sliding surface in~(\ref{surf}) can be constructed to be 
\begin{equation}
	\label{surfp}
	\varLambda  \left( t \right) =s^T\left( t \right) s\left( t \right). 
\end{equation}

For the sake of simplicity, the case that $\bar{x}\left( t \right) \subseteq \bar X_i,i\in \varPhi $ is considered exclusively in this proof. This proof can be extended to the case that $\bar x(t) \subseteq \bar X_0$ similarly.

Then, from~(\ref{surf}) and~(\ref{cont}), we have
\begin{align}
	\label{surfwf}
	\dot{s}\left( t \right) =&S_x\dot{x}\left( t \right) -S_x\left( A_ix\left( t \right) +B_iu\left( t \right) +C_i \right)+S_u\dot{u}\left( t \right) -S_u\left( F_ix\left( t \right) +G_iu\left( t \right) + D_i\right) \notag \\
	=&S_x\left[ \varDelta A_i\left(t\right)x\left( t \right) +\varDelta B_i\left(t\right)u\left( t \right) +\varDelta C_i\left(t\right) \right]-\left( \gamma +\alpha _i +\upsilon _i \left( t \right) \right) \mbox{sgn}\left( s\left( t \right) \right).
\end{align}

Substituting (\ref{surfp}) into (\ref{surfwf}) yields
\begin{equation}
	\dot{\varLambda}\left( t \right) =2s^T\left( t \right) \dot{s}\left( t \right) =2s^T\left( t \right) \left\{ S_x\left[ \varDelta A_i\left(t\right)x\left( t \right) +\varDelta B_i\left(t\right)u\left( t \right) +\varDelta C_i \left(t\right)\right] -\left( \gamma +\alpha _i+\upsilon _i\left( t \right) \right) \mbox{sgn} \left( s\left( t \right) \right) \right\} .
	\label{surfpwf-f} 
\end{equation}

Since we have
\begin{align}
	&2s^T\left( t \right) S_x\left[ \varDelta A_i\left(t\right)x\left( t \right) +\varDelta B_i\left(t\right)u\left( t \right) +\varDelta C_i \left(t\right)\right] \notag \\
	\leqslant &2\left\| s\left( t \right) \right\| \left\| S_x \right\| \left\| \varDelta A_i\left(t\right)x\left( t \right) +\varDelta B_i\left(t\right)u\left( t \right) +\varDelta C_i \left(t\right)\right\| 
	\notag \\
	\leqslant& 2\left\| s\left( t \right) \right\| \left\| S_x \right\| \left( \left\| \varDelta A_i\left(t\right)x\left( t \right) +\varDelta B_i\left(t\right)u\left( t \right) \right\| +\left\| \varDelta C_i\left(t\right) \right\| \right) \notag \\
	\leqslant&2\left\| s\left( t \right) \right\| \left\| S_x \right\| \left( \varepsilon _f\left\| \left[ x^T\left( t \right) ,u^T\left( t \right) \right] ^T \right\| +\varepsilon _g \right) 
	\notag \\
	=&2\varepsilon _f\left\| s\left( t \right) \right\| \left\| S_x \right\| \left\| \left[ x^T\left( t \right) ,u^T\left( t \right) \right] ^T \right\| +2\varepsilon _g\left\| s\left( t \right) \right\| \left\| S_x \right\|
	\label{surfpwf-s} 
\end{align}
and
\begin{equation}
	2\left( \gamma +\alpha _i+\upsilon _i\left( t \right) \right)\left\| s\left( t \right) \right\|\leqslant 2s^T\left( t \right) \left( \gamma +\alpha _i+\upsilon _i\left( t \right) \right) \mbox{sgn} \left( s\left( t \right) \right),
	\label{surfpwf-t} 
\end{equation}
then
\begin{equation}\dot{\varLambda}\left( t \right) \leqslant 2\varepsilon _f\lVert s\left( t \right) \rVert \lVert S_x \rVert \lVert \left[ x^T\left( t \right) , u^T\left( t \right) \right] ^T \rVert+2\varepsilon _g\lVert s\left( t \right) \rVert \lVert S_x \rVert -2\left(\gamma +\alpha  _i +\upsilon _i \left( t \right) \right) \lVert s\left(t\right) \rVert.
	\label{surfpwf-b} 
\end{equation}

Combining~(\ref{errd})-(\ref{surfpwf-b}), one can conclude that
\begin{equation}
	\dot{\varLambda }\left( t \right) \leqslant -2\gamma \lVert s\left( t \right) \rVert =-2\gamma \sqrt{\varLambda  \left( t \right)},
	\label{sliding_re}
\end{equation}
which implies that in finite time, $s \left(t \right) $ can converge to zero. Since initially $s\left(0\right)=0$ and, consequently, the piecewise integral sliding surface in~(\ref{surf}) is maintained subsequently. $\hfill\blacksquare$ 

\begin{remark}
In this paper, we only consider the nominal general nonlinear systems as in (1). Nevertheless, practical nonlinear plants often face issues such as input saturation, undirectional input constraints, dead-zone, and unmodeled dynamics. There have been several control strategies in the literature addressing these issues, such as those in~\cite{li2013adaptive,jiang2018novel,tong2013adaptive,wen2017adaptive,yang2020event}, where neural networks or robust integral terms are included in the control law. It is worth pointing out that (i) sliding-mode control design under control constraints has been a tough research topic and more efforts must be made to solve this problem; (ii) the proposed approach in this paper still works when the dead-zone function can be described by a smooth function, and the unmodeled dynamics have norm-bounds as in (\ref{erro1}); and (iii) the existing approaches, like those in~\cite{li2013adaptive} and~\cite{jiang2018novel} cannot be extended to our case in (\ref{nonl}) because the control input gain cannot be represented by a constant matrix. However, it would be an interesting research topic to investigate the constraint control design problem by integrating the ideas in~\cite{tong2013adaptive,wen2017adaptive,yang2020event} and the proposed ISMPC approach.
\end{remark}

Denote $\bar{A}_i=\left[ A_i,B_i \right], 
	\bar{C}_i=\left[C_i^T, D_i^T \right]^T, \bar{K}_i=\left[F_i, G_i \right], \bar{S} =\left[ S_x, S_u \right], R_1=\left[ I_n,0_{n\times m} \right] ^T, R_2=\left[ 0_{m\times n},I_m \right] ^T.$

Then, one has a more compact form of~(\ref{surf}) as
\begin{gather}
	\textit{for}\,\, \bar{x}\left( t \right) \subseteq \bar X_i, i\in \varphi \notag \\
	s\left( t \right) =\bar{S} \Biggl\{ \bar{x}\left( t \right) -\bar{x}\left(0\right)-\int_0^t{\left[ \left( R_1\bar{A}_i+R_2\bar{K}_i \right) \bar{x}\left(\tau \right) +\bar{C}_i \right] d\tau} \Biggr\}= 0.
	\label{resurf}
\end{gather}

In the literature, the PWA
\begin{gather}
	\emph{for}\,\, \bar{x}\left( t \right) \subseteq \bar X_i, i\in \varphi \notag \\
	\dot{\bar{x}}\left( t \right) =\left( R_1\bar{A}_i+R_2\bar{K}_i \right) \bar{x}\left( t \right) +\bar{C}_i
	\label{nomi}
\end{gather}
is usually named the ``nominal closed-loop control system". One observes that the integral sliding surface variable  $s\left( t \right) $ in~(\ref{resurf}) is in fact the real time difference  between the trajectories of the practical closed-loop control system defined in Remark~\ref{remark3.1} and those of~(\ref{nomi}), multiplying by a weight matrix $\bar S$. This represents the core idea of the proposed ISMPC approach, i.e., to achieve a sliding motion that is as close to (\ref{nomi}) as possible. In other words, one can design~(\ref{nomi}) according to desired control criteria to force the practical closed-loop control system to behave desirable control performance.

The following lemma provides a constructive procedure for
designing~(\ref{nomi}):

\begin{lemma}\cite{rodrigues2005piecewise}
	Given a series of matrices $D_i, i\in \varPhi$ of (\ref{nomi}), which is generated by Algorithm 1,
	the PWA in~(\ref{nomi}) is asymptotically stable, if the following LMIs are feasible w.r.t. a set of scalars $\lambda _i>0$, a matrix $W\in \Re ^{\left( m+n \right) \times \left( m+n \right)}>0$, and matrices $H_j\in \Re ^{m\times \left( m+ n \right)}, j\in \varphi$, and furthermore, the control gains are given as $\bar{K}_j=H_jW^{-1}$:
	\begin{gather}
		\label{lemmaf}
			R_1\bar{A}_0W+R_2H_0+\left( R_1\bar{A}_0W+R_2H_0 \right) ^T<0,\notag\\
				\left[ \begin{matrix}
					\varOmega _i-\lambda _i\bar{C}_i\bar{C}_{i}^{T}&		WQ_{i}^{T}-\lambda _{i}\bar{C}_if_{i}^{T}\\
					\star&		\lambda _{i}\left( I-f_{i}f_{i}^{T} \right)\\
				\end{matrix} \right] <0,
	\end{gather}
	where $\varOmega _i=R_1\bar{A}_iW+R_2H_i+\left( R_1\bar{A}_iW+R_2H_i \right) ^T$ and $Q_i$ and $f_i$ are defined in~(\ref{par1}). 
	\label{lemma1}	
\end{lemma}

\textbf{\emph{Proof.}} Lemma~\ref{lemma1} derives directly from Lemma 4.1 in~\cite{rodrigues2005piecewise} and the proof is omitted here.

The detailed design procedure to calculate the control matrices $\bar{K}_i$ and $D_i$  in~(\ref{nomi}) is summarized as follows:

\noindent\textbf{Algorithm 1} (Sample Method~\cite{rodrigues2005piecewise}):

\emph{Step~1}: Choose a grid for the domain of the vector $\left[ D_1\,\,D_2\,\,\cdots \,\,D_l \right]$ and sample its value at $N$ points.

\emph{Step~2}: Solve Lemma~\ref{lemmaf} for each point in the grid. If a feasible solution is given, stop; 

\emph{Step~3}: Increase the sampling density of the grid and return to Step 1.

\subsection{The Sliding Motion}
\label{sec:SASM} How to obtain the sliding surface matrix $\bar S$ through stability analysis of the sliding motion will be further shown in this subsection. For the sliding motion w.r.t.~(\ref{surf}), two conditions must be fulfilled simultaneously:
\begin{align}
	s\left( t \right) = & 0, \notag \\
	\dot{s}\left( t \right) =&S_x\dot{x}\left( t \right) -S_x\left( A_ix\left( t \right) +B_iu\left( t \right) +C_i \right)+S_u\dot{u}\left( t \right) -S_u\left( F_ix\left( t \right) +G_iu\left( t \right)+D_i \right) =0.
	\label{condis}
\end{align}

Since $S_u$ is nonsingular, (\ref{condis}) yields
\begin{gather}
	\textit{for}\ \bar{x}\left( t \right) \subseteq \bar X_i, i\in \varphi \notag \\
	\dot{u}\left( t \right) =F_ix\left( t \right) +G_iu\left( t \right) +D_i-S_{u}^{-1}S_x\left[  \varDelta A_i\left(t\right)x\left( t \right) +\varDelta B_i\left(t\right)u\left( t \right)  +\varDelta C_i \left(t\right)\right],
	\label{eqdy}
\end{gather}
which is usually named the \emph{equivalent sliding-mode control law}. The corresponding sliding motion is then referred to the dynamical
system consisting of~(\ref{pls}) and~(\ref{eqdy}):
\begin{gather}
	\textit{for}\,\, \bar{x}\left( t \right) \subseteq X_i, i\in \varphi \notag \\
	\left\{ \begin{aligned}
		\dot{x}(t)=&A_ix(t)+B_iu(t)+C_i+\varDelta A_i\left(t\right)x(t)+\varDelta B_i\left(t\right)u(t)+\varDelta C_i\left(t\right)\\
		\dot{u}(t)=&F_ix(t)+G_iu(t)+D_i-S_{u}^{-1}S_x\left[ \varDelta A_i\left(t\right)x(t)+\varDelta B_i\left(t\right)u(t)+\varDelta C_i\left(t\right) \right]\\
	\end{aligned} \right. 
	\label{clody}
\end{gather}
or equivalently
\begin{gather}
	\textit{for}\,\, \bar{x}\left( t \right) \subseteq \bar X_i, i\in \varphi \notag \\
	\dot{\bar{x}}\left( t \right) =\left( R_1\bar{A}_i+R_2\bar{K}_i+\left( R_1-R_2S_{u}^{-1}S_x \right) \varDelta \bar{A}_i\left(t\right) \right) \bar{x}\left( t \right) +\bar{C} _i+\left( R_1-R_2S_{u}^{-1}S_x \right) \varDelta C_i\left(t\right).
	\label{eqslm}
\end{gather}

One has the following theorem:

\begin{theorem}
	The asymptotic stability of the sliding motion in~(\ref{eqslm}) can be ensured, when the LMIs in~(\ref{lmi2}) are feasible w.r.t. a series of positive scalars $\eta _0$, $\eta _{i_1}, \eta _{i_2}, \eta _{i_3}, i\in \varPhi $ and a matrix $P\in \Re ^{(m+n) \times (m+n)}>0$. Additionally, the integral sliding surface matrix is $\bar{S}=R_{2}^{T}P$:
	\begin{gather}
			\left[ \begin{matrix}
				\varLambda _0+\eta _0\varepsilon _{f_0}^{2}I_{m+n}&		PR_1&		PR_2\\
				\star&		R_{1}^{T}PR_1-\eta _0I_n&		0\\
				\star&		\star&		-R_{2}^{T}PR_2\\
			\end{matrix} \right] <0,\notag\\
		\left[ \begin{matrix}
			\varLambda _i+\eta _{i_1}\varepsilon _{f}^{2}I_{m+n}-\eta _{i_3}Q_{i}^{T}Q_i&		PR_1&		PR_1&		P\bar{C}_i-\eta _{i_3}Q_{i}^{T}f_i&		PR_2\\
			\star&		R_{1}^{T}PR_1-\eta _{i_1}I_n&		0&		0&		0\\
			\star&		\star&		R_{1}^{T}PR_1-\eta _{i_2}I_n&		0&		0\\
			\star&		\star&		\star&		\eta _{i_2}\varepsilon _{g}^{2}-\eta _{i_3}\left( f_{i}^{\text{T}}f_i-1 \right)&		0\\
			\star&		\star&		\star&		\star&		-\frac{1}{2}R_{2}^{T}PR_2\\
		\end{matrix} \right] <0, i\in \varPhi,
			\label{lmi2}
	\end{gather}
	where $\varLambda _i=P\left( R_1\bar{A}_i+R_2\bar{K}_i \right) +\left( R_1\bar{A}_i+R_2\bar{K}_i \right) ^TP, i\in \varphi.$
	\label{thm2}
\end{theorem}

\textbf{\emph{Proof}}:~The following Lyapunov function is used
for stability analysis:
\begin{equation}
	\label{lyp}
	V\left( \bar{x}\left( t \right) \right) =\bar{x}^T\left( t \right) P\bar{x}\left( t \right). 
\end{equation}

Along the trajectories of the sliding motion in~(\ref{eqslm}), the derivative of (\ref{lyp}) can be obtained as
\begin{align}
	&\dot{V}\left( \bar{x}\left( t \right) \right) =\dot{\bar{x}}^T\left( t \right) P\bar{x}\left( t \right) +\bar{x}^T\left( t \right) P\dot{\bar{x}}\left( t \right) 
	\notag \\
	&=\dot{\bar{x}}^T\left( t \right) \left. \Bigl\{ P\left( R_1\bar{A}_i+R_2\bar{K}_i \right) +\left( R_1\bar{A}_i+R_2\bar{K}_i \right) ^TP \right. +\varDelta \bar{A}_{i}^{T}\left(t\right)\left( PR_1-PR_2S_{u}^{-1}S_x \right) ^T+\left. \left( PR_1-PR_2S_{u}^{-1}S_x \right) \varDelta \bar{A}_i\left(t\right) \right. \Bigr\} \bar{x}\left( t \right) 
	\notag \\
	&+\bar{x}^T\left( t \right) P\bar{C}_i+\bar{C}_{i}^{T}P\bar{x}\left( t \right) +\bar{x}^T\left( t \right) PR_1\varDelta C_i\left(t\right)+\varDelta C_{i}^{T}\left(t\right)R_{1}^{T}P\bar{x}\left( t \right) -\bar{x}^T\left( t \right) PR_2S_{u}^{-1}S_x\varDelta C_i\left(t\right)
	-\left(\bar{x}^T\left( t \right) PR_2S_{u}^{-1}S_x\varDelta C_i \left(t\right)\right) ^T.
	\label{lypwf}
\end{align}

Let $M$ be a positive definite matrix satisfying $M=\sqrt{P}$. Then, it follows from the fact $\bar{S}=R_{2}^{T}P$ that
\begin{align}
	\label{lemma11}
	&-PR_2S_{u}^{-1}S_x\varDelta \bar{A}_i\left(t\right)-\left( PR_2S_{u}^{-1}S_x\varDelta \bar{A}_i\left(t\right) \right) ^T \notag
	\\
	=&-PR_2\left( R_{2}^{T}PR_2 \right) ^{-1}R_{2}^{T}MMR_1\varDelta \bar{A}_i\left(t\right)-\left(PR_2\left( R_{2}^{T}PR_2 \right) ^{-1}R_{2}^{T}MM R_1\varDelta \bar{A}_i\left(t\right) \right) ^T \notag
	\\
	\leqslant& PR_2\left( R_{2}^{T}PR_2 \right) ^{-1}R_{2}^{T}MM R_2\left( R_{2}^{T}PR_2 \right) ^{-1}R_{2}^{T}P
	+\varDelta \bar{A}_{i}^{T}\left(t\right)R_{1}^{T}MM R_1\varDelta \bar{A}_i \left(t\right)\notag
	\\
	=&PR_2\left( R_{2}^{T}PR_2 \right) ^{-1}R_{2}^{T}P+\varDelta \bar{A}_{i}^{T}\left(t\right)R_{1}^{T}PR_1\varDelta \bar{A}_i\left(t\right)
\end{align}
and
\begin{align}
	\label{lemma12}
	&-\bar{x}^T\left( t \right) PR_2S_{u}^{-1}S_x\varDelta C_i\left(t\right)-\left( \bar{x}^T\left( t \right) PR_2S_{u}^{-1}S_x\varDelta C_i\left(t\right) \right) ^T \notag
	\\
	=&-\bar{x}^T\left( t \right) PR_2\left( R_{2}^{T}PR_2 \right) ^{-1}R_{2}^{T}MMR_1\varDelta C_i\left(t\right) -\left(\bar{x}^T\left( t \right) PR_2\left( R_{2}^{T}PR_2 \right) ^{-1}R_{2}^{T}MMR_1\varDelta C_i\left(t\right) \right) ^T \notag
	\\
	\leqslant& \bar{x}^T\left( t \right) PR_2\left( R_{2}^{T}PR_2 \right) ^{-1}R_{2}^{T}MMR_2\left( R_{2}^{T}PR_2 \right) ^{-1}R_{2}^{T}P\bar{x}\left( t \right) +\varDelta C_i^{T}\left(t\right)R_{1}^{T}MMR_1\varDelta C_i\left(t\right) \notag
	\\
	=&\bar{x}^T\left( t \right) PR_2\left( R_{2}^{T}PR_2 \right) ^{-1}R_{2}^{T}P\bar{x}\left( t \right) +\varDelta C_i^{T}\left(t\right)R_{1}^{T}PR_1\varDelta C_i\left(t\right).
\end{align}

Then, combining~(\ref{lemma11}) and~(\ref{lemma12}), we can obtain
\begin{align}
	&\dot{V}\left( \bar{x}\left( t \right) \right) \leqslant \bar{x}^T\left( t \right) \left. \Bigl\{\right. \varLambda _i + PR_1\varDelta \bar{A}_i\left(t\right)+\left( PR_1\varDelta \bar{A}_i \left(t\right)\right) ^T
	+\varDelta \bar{A}_{i}^{T}R_{1}^{T}PR_1\varDelta \bar{A}_i\left. +2PR_2\left( R_{2}^{T}PR_2 \right) ^{-1}R_{2}^{T}P \right. \Bigr\} \bar{x}\left( t \right) 
	\notag \\
	+&\bar{x}^T\left( t \right) P\bar{C}_i+\bar{C}_{i}^{T}P\bar{x}\left( t \right) +\varDelta C_{i}^{T}\left(t\right)R_{1}^{T}PR_1\varDelta C_i\left(t\right)
	+\bar{x}^T\left( t \right) PR_1\varDelta C_i\left(t\right)+\varDelta C_{i}^{T}\left(t\right)R_{1}^{T}P\bar{x}\left( t \right). 
	\label{lypxy}
\end{align}

It follows from (\ref{lypxy}) that $\dot{V}\left( \bar{x}\left( t \right) \right) <0$ if
\begin{align}
	&\bar{x}^T\left( t \right) \left. \Bigl\{ 2PR_2\left( R_{2}^{T}PR_2 \right) ^{-1}R_{2}^{T}P+ \right. \varLambda _i+\varDelta \bar{A}_{i}^{T}\left(t\right)R_{1}^{T}PR_1\varDelta \bar{A}_i\left(t\right)\left. +PR_1\varDelta \bar{A}_i\left(t\right)+\left( PR_1\varDelta \bar{A}_i\left(t\right) \right) ^T \right. \Bigr\} \bar{x}\left( t \right) 
	\notag \\
	&+\varDelta C_{i}^{T}\left(t\right)R_{1}^{T}PR_1\varDelta C_i\left(t\right)+\bar{x}^T\left( t \right) P\bar{C}_i+\bar{C}_{i}^{T}P\bar{x}\left( t \right)
	+\bar{x}^T\left( t \right) PR_1\varDelta C_i\left(t\right)+\varDelta C_{i}^{T}\left(t\right)R_{1}^{T}P\bar{x}\left( t \right) <0.
	\label{lypbds}
\end{align}

Besides, (\ref{lypbds}) is equivalent to
\begin{equation}
	\left[ \begin{array}{c}
		\bar{x}\left( t \right)\\
		\varDelta \bar{A}_i\left(t\right)\bar{x}\left( t \right)\\
		\varDelta C_i\left(t\right)\\
		1\\
	\end{array} \right] ^T	\left[ \begin{matrix}
		\varLambda _i+\varPi _0&		PR_1&		PR_1&		P\bar{C}_i\\
		\star&		R_{1}^{T}PR_1&		0&		0\\
		\star&		\star&		R_{1}^{T}PR_1&		0\\
		\star&		\star&		\star&		0\\
	\end{matrix} \right] \left[ \begin{array}{c}
		\bar{x}\left( t \right)\\
		\varDelta \bar{A}_i\left(t\right)\bar{x}\left( t \right)\\
		\varDelta C_i\left(t\right)\\
		1\\
	\end{array} \right] <0,
	\label{lyplmi}
\end{equation}
where $\varPi _0=2PR_2\left( R_{2}^{T}PR_2 \right) ^{-1}R_{2}^{T}P.$

Furthermore, the norm-bound of approximation errors defined in~(\ref{erro1}) satisfies
\begin{equation}
	\label{errlmi1}
		\left[ \begin{array}{c}
			\bar{x}\left( t \right)\\
			\varDelta \bar{A}_i\left(t\right)\bar{x}\left( t \right)\\
			\varDelta C_i\left(t\right)\\
			1\\
		\end{array} \right] ^T\left[ \begin{matrix}
			-\varepsilon _{f}^{2}I_{m+n}&		0&		0&		0\\
			\star&		I_n&		0&		0\\
			\star&		\star&		0&		0\\
			\star&		\star&		\star&		0\\
		\end{matrix} \right] \left[ \begin{array}{c}
			\bar{x}\left( t \right)\\
			\varDelta \bar{A}_i\left(t\right)\bar{x}\left( t \right)\\
			\varDelta C_i\left(t\right)\\
			1\\
		\end{array} \right] \leqslant 0
\end{equation}
and
\begin{equation}
	\label{errlmi2}
		\left[ \begin{array}{c}
			\bar{x}\left( t \right)\\
			\varDelta \bar{A}_i\left(t\right)\bar{x}\left( t \right)\\
			\varDelta C_i\left(t\right)\\
			1\\
		\end{array} \right] ^T\left[ \begin{matrix}
			0&		0&		0&		0\\
			\star&		0&		0&		0\\
			\star&		\star&		-\varepsilon _{g}^{2}I_n&		0\\
			\star&		\star&		\star&		1\\
		\end{matrix} \right] \left[ \begin{array}{c}
			\bar{x}\left( t \right)\\
			\varDelta \bar{A}_i\left(t\right)\bar{x}\left( t \right)\\
			\varDelta C_i\left(t\right)\\
			1\\
		\end{array} \right] \leqslant 0.
\end{equation}

For the partitions $\bar X_i, i \in \varPhi$, by defining $\xi \left( t \right) =\left[ \begin{matrix}
	\bar{x}^T\left( t \right)&		\left( \varDelta \bar{A}_i\left(t\right)\bar{x}\left( t \right) \right) ^T&		\varDelta C_{i}^{T}&		1\\
\end{matrix} \right] ^T$,~(\ref{par4}) is rewritten as
\begin{equation}
	\label{afflmi}
		\xi^T \left( t \right)\left[ \begin{matrix}
			Q_{i}^{T}Q_i&		0&		0&		Q_{i}^{T}f_i\\
			\star&		0&		0&		0\\
			\star&		\star&		0&		0\\
			\star&		\star&		\star&		f_{i}^{\text{T}}f_i-1\\
		\end{matrix} \right] \xi \left( t \right) \leqslant 0.
\end{equation}

By applying Lemma~\ref{lemmaapen} in Appendix. \ref{sec:a} and combining the LMIs in~(\ref{lyplmi})-(\ref{afflmi}), one can conclude that~(\ref{lyplmi}) is fulfilled if there exists a series of positive scalars $\eta _{i1},\eta _{i2},\eta _{i3},  i\in \varPhi$ such that 
\begin{equation}
	\label{schur}
		\left[ \begin{matrix}
			\varLambda _i+\varPi _0+\eta _{i_1}\varepsilon _{f}^{2}I_{m+n}-\eta _{i_3}Q_{i}^{T}Q_i&		PR_1&		PR_1&		P\bar{C}_i-\eta _{i_3}Q_{i}^{T}f_i\\
			\star&		R_{1}^{T}PR_1-\eta _{i_1}I_n&		0&		0\\
			\star&		\star&		R_{1}^{T}PR_1-\eta _{i_2}I_n&		0\\
			\star&		\star&		\star&		\eta _{i_2}\varepsilon _{g}^{2}-\eta _{i_3}\left( f_{i}^{\mathrm{T}}f_i-1 \right)\\
		\end{matrix} \right] <0.
\end{equation}

Based on the Schur’s complement in~\cite{boyd1994linear}, the LMIs
in~(\ref{schur}) are proved to be equivalent to those in~(\ref{lmi2}). One can obtain that the derivative of  $V\left( \bar{x}\left( t \right) \right) $ is strictly negative if the LMIs in~Theorem \ref{thm2} can hold simultaneously. Therefore, the sliding motion~(\ref{eqslm}) is asymptotically stable if the LMIs in~(\ref{lmi2})  can be fulfilled.$\hfill\blacksquare$ 


It's worth noting that the sliding surface matrix $S_u$ in~(\ref{surf}) is required to be nonsingular because its inverse is essential in constructing the sliding mode controller in~(\ref{cont}). By designing $\bar{S}=R_{2}^{T}P$, one has  $S_u=R_{2}^{T}PR_{2}$ with $R_2=\left[ 0_{m\times n},I_m \right] ^T$. One can then conclude that $S_u$ is positive definite and thus invertible since $P$ is positive definite.

\begin{remark}
	The common quadratic Lyapunov function used as in~(\ref{lyp}) tends to be conservative, especially when dealing with more complicated nonlinear systems. In order to achieve less conservative control synthesis, the more relaxing piecewise/fuzzy Lyapunov functions in~\cite{zhang2017event,wang2020a,farbood2021fuzzy} could be potentially employed. However, preliminary research along this direction has shown that the control design would tend to be extreme complex in practice~\cite{xi2010piecewise}. 
\end{remark}

\begin{corollary}
	If the conditions in Theorem~\ref{thm2}
	are fulfilled simultaneously, the practical closed-loop control system consisting
	of~(\ref{nonl}) and~(\ref{cont}) behaves a semi-globally asymptotically stable
	sliding motion since initially.
	\label{cora3.1}
\end{corollary}

\textbf{\emph{Proof}}:~Corollary~\ref{cora3.1} can be concluded based on the fact
that only within $X \times U$, the PALM in~(\ref{pls}) is equivalent to the original nonlinear system in~(\ref{nonl}). The proof will be thus omitted here.

A systematic algorithm for implementing the ISMPC approach concerning the nonlinear systems in~(\ref{nonl}) can be summarized as follows:

\noindent\textbf{Algorithm 2:} For any controlled nonlinear system in~(\ref{nonl}), a  piecewise sliding-mode parallel controller in~(\ref{cont}) can be constructed such that a semi-globally asymptotically stable sliding
motion can be achieved since initially, by conducting the subsequent procedure:

\emph{Step 1}: To obtain the PALM via the linearization approach of the controlled nonlinear system $X \times U$.

\emph{Step 2}: To obtain the control matrices $\bar{K}_i=\left[ F_i,G_i \right]$ and $D_i$ via Algorithm 1.

\emph{Step 3}: To obtain the sliding surface matrix $\bar{S}$ based on Theorem~\ref{thm2}.

\emph{Step 4}: To choose a suitable parameter $\gamma>0$ to ensure that the system trajectories move
within $X \times U$. In case  Algorithm~2 can not return a feasible solution, raise the amount of partitions $l$ and repeat the loop until $l$ exceeds the pre-chosen threshold.

\begin{remark}
The computational complexity of Algorithm 2 consists of implementing Algorithm 1 and solving the  LMIs~(\ref{lmi2}). Following the result~\cite{wang2014outage},  given the maximum amount of sampled points in the grid $M$, the complexity of Algorithm 1 could be  calculated as $\mathcal{O} \left(Ml^{\frac{3}{2}}n^{\frac{9}{2}} \right)$; similarity, the complexity of solving the LMIs~(\ref{lmi2}) is $\mathcal{O} \left(l^{\frac{3}{2}}n^{\frac{9}{2}} \right)$. Therefore, the computational complexity of Algorithm 2 is $\mathcal{O} \left(Ml^{\frac{3}{2}}n^{\frac{9}{2}} \right)$.
\end{remark}

\subsection{Universality Discussion}
\label{sec:UPCC}
In Subsections~\ref{sec:DISS} and~\ref{sec:SASM}, we have developed an ISMPC scheme to stabilize a general nonlinear system as in~(\ref{nonl}) through PALMs. It is shown that the scheme works if a series of LMIs is fulfilled simultaneously. Therefore, the universality of such a scheme is questionable, which motivates the study in the remaining of this section. To be specific, we will answer the subsequent question: for any given
stabilizable nonlinear system in~(\ref{nonl}), can one always design a piecewise sliding-mode parallel controller as in~(\ref{cont}) such that the resultant closed-loop control system behaves a
stable sliding motion since initially?

In the remaining of this paper, we say that a general non-affine continuous-time nonlinear
system in (\ref{nonl}) is GAS/GES, if there exists a control law
in the form of $\dot{u}\left( t \right) =g\left( x\left( t \right) ,u\left( t \right) \right)$ such that
\begin{align}
	\label{clcs}
	\left\{ \begin{aligned}
		\dot{x}\left( t \right) &=f\left( x\left( t \right) ,u\left( t \right) \right)\\
		\dot{u}\left( t \right) &=g\left( x\left( t \right) ,u\left( t \right) \right)\\
	\end{aligned} \right.  
\end{align}
is globally asymptotically/exponentially stable. For brevity, we
rewrite~(\ref{clcs}) as
\begin{equation}
	\dot{\bar{x}}\left( t \right) =F\left( \bar{x}\left( t \right) \right) =R_1f\left( x\left( t \right) ,u\left( t \right) \right) +R_2g\left( x\left( t \right) ,u\left( t \right) \right)
	\label{nonre}
\end{equation}
where $R_1$ and $R_2$ are defined in Section \ref{sec:DISS} below~(\ref{sliding_re}). 

The subsequent two theorems (Theorems~\ref{thm:GES} and~\ref{thm:GAS}) summarize the main results of this subsection.

\begin{theorem} 
	For a GES nonlinear system in~(\ref{nonl}), one can always design a piecewise sliding-mode parallel
	controller as in (\ref{cont}) such that the resultant closed-loop control
	system behaves a semi-globally exponentially stable sliding motion
	from the beginning of evolution.
	\label{thm:GES}
\end{theorem}

\textbf{\emph{Proof.}} See Appendix. \ref{proof:GES}.

Before proceeding with the more general case,  a preliminary result from the Lyapunov converse theorem in~\cite{khalil2002nonlinear} is presented.

Suppose the nonlinear system in~(\ref{nonre}) is globally asymptotically stable, then  along the trajectories of (\ref{nonre}), there exist two functions $\alpha _1\left( \cdot \right) $ and $\alpha _2\left( \cdot \right) $ belonging to the class $\mathcal{K} _\infty$, a  function $\alpha _3\left( \cdot \right) $ belonging to the class $\mathcal{K}$, a scalar $h>0$, and a Lyapunov function
$V\left( \bar{x} \right)$ satisfying
\begin{align}
	\label{slyp1}
	\alpha _1\left( \lVert \bar{x} \rVert \right) \leqslant V\left( \bar{x} \right) &\leqslant \alpha _2\left( \lVert \bar{x} \rVert \right)
	\\
	\label{slyp2}
	\frac{\partial V\left( \bar{x} \right)}{\partial \bar{x}}F\left( \bar{x} \right) &\leqslant -\alpha _3\left( \lVert \bar{x} \rVert \right)
	\\
	\label{slyp3}
	\lVert \frac{\partial V\left( \bar{x} \right)}{\partial \bar{x}} \rVert &\leqslant h
\end{align}
where $F\left( \bar{x} \right)$ is defined in~(\ref{nonre}).

\begin{theorem}
	\label{thm:GAS}
	For a GAS nonlinear system in~(\ref{nonl}), one can
	always design a piecewise  sliding-mode parallel
	controller as in~(\ref{cont}) such that the resultant closed-loop control
	system behaves a semi-globally asymptotically stable sliding motion from the beginning of evolution, if for the function $\alpha_3\left( \cdot \right)$ defined in~(\ref{slyp2}), the condition
	\begin{equation}
		\label{thm5.1}
		\inf_{\bar{x} \in X\times U} 
		\alpha _3\left( \lVert \bar{x}\rVert \right)\geqslant \rho \lVert \bar{x} \rVert+\alpha _4\left( \lVert \bar{x}\rVert \right)
	\end{equation}
	holds for a $\mathcal{K}$ function
	$\alpha _4\left( \cdot \right) $ and a scalar $\rho>0$.
\end{theorem}

\textbf{\emph{Proof.}} See Appendix.~\ref{proof:GAS}.


\begin{remark}
	As shown in Appendix. \ref{proof:GES} and Appendix. \ref{proof:GAS},  the construction of the piecewise sliding-mode parallel control law depends on the norm-bounds of the uncertainties. Therefore, for a given GES nonlinear system or a GAS nonlinear system satisfying the condition in~(\ref{thm5.1}), one can always design a corresponding parallel control law to stabilize the original system in~(\ref{nonl}) by decreasing the norm-bounds of uncertainties until the LMIs in Theorem~\ref{thm2} are fulfilled. This fact brings us great confidence in applying the easy-checking control design approach in~Theorem~\ref{thm2} to industrial practice.
\end{remark}

\section{Simulation Studies}
\label{sec:sims}
Two different types of numerical examples are given to demonstrate the effectiveness and advantages of the developed ISMPC approach.

\subsection{Nonlinear Chua’s Circuit}
The famous Chua’s circuit has the following
dynamical equation~\cite{chua1986canonical}:
\begin{align*}
	C_1\frac{dx_1}{dt}&=\frac{x_2-x_1}{R}-g\left( x_1 \right) -u\\
	C_2\frac{dx_2}{dt}&=\frac{x_1-x_2}{R}-x_3\\
	L\frac{dx_3}{dt}&=x_2-v_d
\end{align*}
where $x_1$ and $x_2$ are the voltages across the capacitors $C_1$ and $C_2$, respectively,
$x_3$ is the current passing the inductor, $u$ is the control current used to stabilize the nonlinear circuit and $v_d$ is the voltage loss $R_0x_3$ or the external disturbance. The nonlinear function $g\left(x_1\right)$ describes the nonlinearity of the resistor is given by
\begin{align*}
	g\left( x_1 \right) =ax_1+cx_{1}^{3},a<0,c>0.
\end{align*}


Here we use the approximation model built in~\cite{zhang2007output}, where the system parameters and subspaces can be referred to, for the control design. Note that the dynamical equation of the Chua's circuit could be reformulated as $\dot{x}\left(t\right)=f(x\left(t\right))+Bu\left(t\right)$, where $f(x\left( t \right) )=\left[ \begin{matrix}
	-\frac{g\left(x_1\right)}{C_1}&		0&		0\\
\end{matrix} \right] ^T+\left[ \begin{matrix}
	-\frac{1}{C_1R}&		\frac{1}{C_1R}&		0\\
	\frac{1}{C_2R}&		-\frac{1}{C_2R}&		-1\\
	0&		\frac{1}{L}&		0\\
\end{matrix} \right] x\left(t\right)$ and $B=\left[ \begin{matrix}
	-\frac{1}{C_1}&		0&		0\\
\end{matrix} \right] ^T$. 

\begin{figure}[htb]
	\centering
	\subfigure[The voltages across the capacitors]{
		\begin{minipage}[t]{0.48\textwidth}
			\centering
			\includegraphics[width=\textwidth]{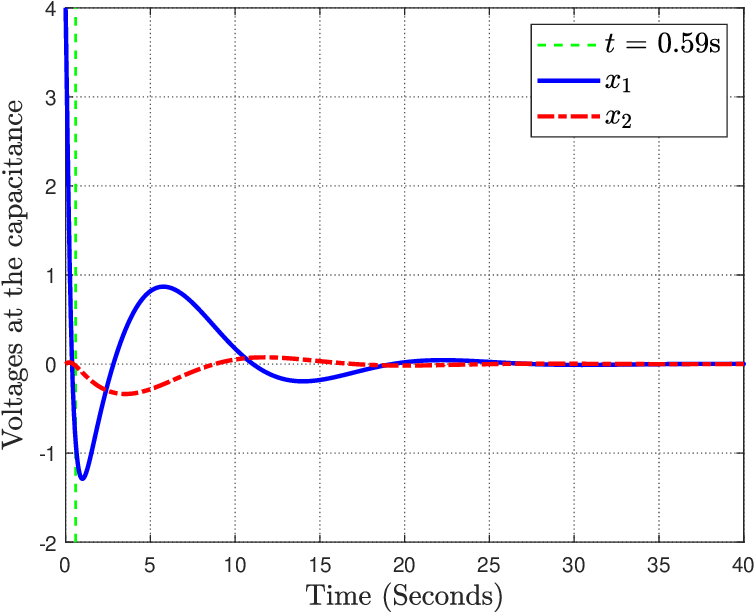}
		\end{minipage}	
	}
	\subfigure[The currents trajectories]{
		\begin{minipage}[t]{0.48\textwidth}
			\centering
			\includegraphics[width=\textwidth]{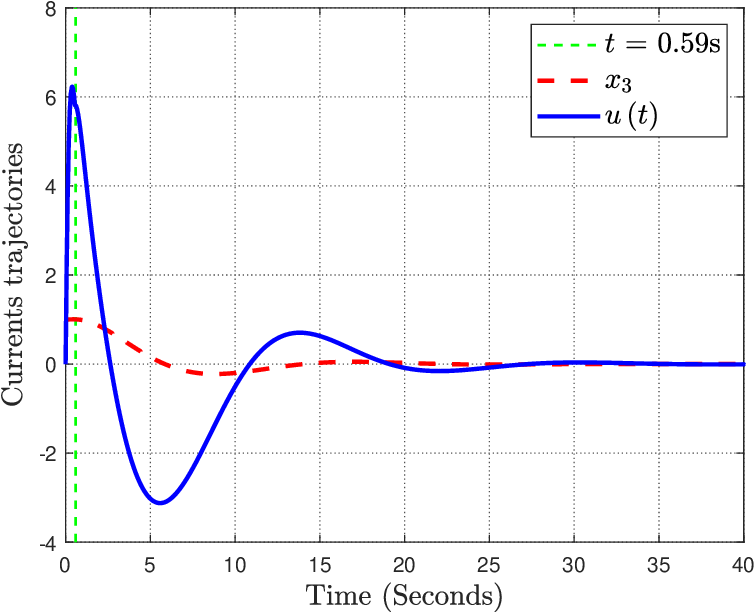}
		\end{minipage}	
	}
	\subfigure[The integral sliding-mode function]{
		\begin{minipage}[t]{0.48\textwidth}
			\centering
			\includegraphics[width=\textwidth]{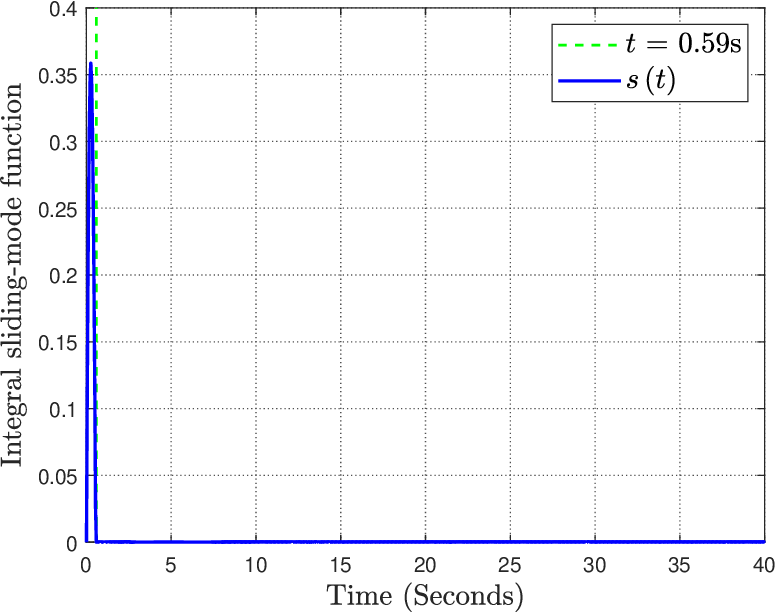}
		\end{minipage}	
	}
	\caption{The simulation results}
	\label{ex:state}
\end{figure}

In practice, it is very difficult to calculate the approximation error bounds precisely. In this experiment, following the approach as in~\cite{gao2013universal}, the norm-bounds of uncertainties are determined by 
\begin{align}
\varepsilon _{f_0}=&\max \left\{ \frac{\left\| f(x\left( t \right) )-A_0x\left( t \right) \right\|}{\left\| x\left( t \right) \right\|} \right\}, \varepsilon _f=\frac{1}{2}\max \left\{ \frac{\left\| A_ix\left( t \right) +C_i-f(x\left( t \right) ) \right\|}{\left\| x\left( t \right) \right\|} \right\} ,\notag\\ 
\varepsilon _g=&\frac{1}{2}\max \left\| A_ix\left( t \right) +C_i-f(x\left( t \right) ) \right\|, i =\left\{1,2\right\}
\label{norm11}
\end{align}
at a series of  vertex points, which could be sampled uniformly or randomly,
within the operating region $\left(x_{1}, x_{2},x_{3}\right) \in\left[ -5,5 \right] \times \left[ -5,5 \right] \times \left[ -5,5 \right] $, which yields $\varepsilon _{f_0}=0.007$, $\varepsilon _{f}=0.003$, and $\varepsilon _{g}=0.005$ by numerically calculating~(\ref{norm11}) on these sampled vertex points. It is noticed that only finite points could be implemented. However, one could sample more vertex points within the operating  region to enhance the precision of the obtained norm-bounds of uncertainties.

Based on Algorithm 2, one can obtain the controller matrices and the integral sliding surface matrix as
\begin{align*}
	K_0 &= \left[ \begin{matrix}
		6.0518&		49.6777&		20.8074&		-2.5596\\
	\end{matrix} \right], 
	K_1 = \left[ \begin{matrix}
		6.1412&		49.5742&		20.8064&		-2.5411\\
	\end{matrix} \right], \\
	K_2 &= \left[ \begin{matrix}
		5.7869&		48.5033&		20.3217&		-2.5140\\
	\end{matrix} \right],
	D_1 =-D_2=0.200, \\
	\bar{S} &=\left[ \begin{matrix}
		-0.3318&		-4.8582&		-1.6437&		0.4322\\
	\end{matrix} \right] .
\end{align*}

\emph{In particular, the PALM of the Chua's circuit possesses an identical constant input matrix in each subspace.} It can be observed that the Chua's circuit behaves desirable control performance.

In order to avoid singular problem and reduce chattering phenomenon,  $\mbox{sgn}\left( s\left( t \right) \right) $ in~(\ref{cont}) is replaced by its approximation function:
	$$\frac{s\left( t \right)}{\lVert s\left( t \right) \rVert +\delta  }$$
with $\delta = 0.001$.

This simulation sets $\left[ \begin{matrix}
	4&		0&		1&		0\\
\end{matrix} \right] $ as the initial state $\bar{x}\left(0\right)$. The voltages across the capacitors $C_1$ and $C_2$ are shown in Fig.~\ref{ex:state}(a), respectively, and the currents in the circuit are presented in Fig.~\ref{ex:state}(b). One can observe that the control input is pretty smooth during the simulation time.

It is also observed from  Fig.~\ref{ex:state}(c)  that the integral sliding surface can be reached and maintained since the time $t=0.59\rm s$, which is prior to the time when the system trajectories converge to zero (which is about $t=25\rm s$ according to Fig.~\ref{ex:state}(a) and Fig.~\ref{ex:state}(b). This coincides with the theoretical analysis in this paper that the controlled system should enter the sliding mode first and then behave stable sliding motion. Note that theoretically, the practical closed-loop control system should enter and keep the sliding mode from the beginning of this simulation.
The approximation signum function utilized in the constructed controller in~(\ref{cont}) results in this deviation from the ideal sliding mode. Note that the approximation signum function is used to avoid critical chattering phenomenon during the sliding motion. Better approximation can be achieved if the positive constant $\delta $ is chosen to be smaller, which yields better control performance (closer to the ideal case). However, more evident chattering phenomenon would be caused as a result, which is undesirable in practice. How to determine the value of $\delta$ is a tradeoff in practice and depends on specific applications.

\subsection{Inverted Pendulum}
Stabilization of the inverted pendulum is always used to demonstrate the advantages and effectiveness of various control methods. The inverted pendulum system in~\cite{gao2013universal} is chosen. The inverted pendulum has the following dynamics:
\begin{align*}
	\dot{x}_1\left( t \right) &=x_2\left( t \right)\\
	\dot{x}_2\left( t \right) &=\frac{g\sin \left( x_1 \right) -amlx_{2}^{2}\sin \left( 2x_1 \right) /2-a\cos \left( x_1 \right) u(t)}{4l/3-aml\cos ^2\left( x_1 \right)}
	\label{inp}
\end{align*}
where $x_1$ is the angle of pendulum from the vertical, $x_2$ denotes the angular velocity, and $u(t)$ is the input signal.  $g=9.8\rm m/s^2$ is called as the gravity constant, $M=4.0\rm kg$ is the mass of the cart, the mass and length of the pendulum is $m=2.0\rm kg$ and $l=0.5\rm m$ respectively, $a=\frac{1}{M+m}$.


$X\times U$ is selected as $\left[ -\frac{\pi}{2},\frac{\pi}{2} \right] \times \left[ -3,3 \right] \times \left[ -300,300 \right] $. To obtain the corresponding PLAM of the inverted pendulum by linearization around the operating points $\left( 0;0;0 \right) , \left( \pm \frac{\pi}{3};0;0 \right) $ and $\left( \pm \frac{13\pi}{30};0;0 \right) $. And the subspaces are selected as 
\begin{align*}
	\mathcal{P}_0&=\left\{\bar{x}|-\frac{\pi}{3}\leqslant
	x_1\leqslant \frac{\pi}{3} \right\},
	\mathcal{P}_1=\left\{\bar{x}|\frac{\pi}{3}\leqslant
	x_1\leqslant \frac{5\pi}{12} \right\},
	\mathcal{P}_2=\left\{\bar{x}|\frac{5\pi}{12}\leqslant
	x_1\leqslant \frac{\pi}{2} \right\},\\
	\mathcal{P}_3&=\left\{\bar{x}|-\frac{5\pi}{12} \leqslant
	x_1\leqslant -\frac{\pi}{3}\right\},
	\mathcal{P}_4=\left\{\bar{x}|-\frac{\pi}{2} \leqslant
	x_1\leqslant -\frac{5\pi}{12} \right\}.
\end{align*}

One obtains
\begin{align*}
	\bar{A}_0=&\left[ \begin{matrix}
		0&		1&		0\\
		19.6000&		0&		-0.6667\\
	\end{matrix} \right] 
	,
	\bar{A}_1=\bar{A}_3=\left[ \begin{matrix}
		0&		1&		0\\
		4.7040&		0&		-0.2667\\
	\end{matrix} \right],  \bar{A}_2=\bar{A}_4=\left[ \begin{matrix}
		0&		1&		0\\
		1.5955&		0&		-0.1585\\
	\end{matrix} \right],\\
	C_0=&\left[ \begin{array}{c}
		0\\
		0\\
	\end{array} \right] , C_1=-C_3=\left[ \begin{array}{c}
		0\\
		8.6533\\
	\end{array} \right],
	C_2=-C_4=\left[ \begin{array}{c}
		0\\
		12.3638\\
	\end{array} \right].  
\end{align*}

Notice that the dynamics equation of the inverted pendulum can be expressed as an affine nonlinear form by $\dot{x}\left(t\right)=f(x\left(t\right))+g(x\left(t\right))u\left(t\right)$, where $f(x\left( t \right) )=\frac{g\sin x_1-amlx_{2}^{2}\sin \left( 2x_1 \right) /2}{4l/3-aml\cos ^2x_1}$ and $g(x\left( t \right) )=\frac{-a\cos x_1}{4l/3-aml\cos ^2x_1}$. 

\begin{figure}[hbt]
	\centering
	\subfigure[The system state]{
		\begin{minipage}[t]{0.46\textwidth}
			\centering
			\includegraphics[width=\textwidth]{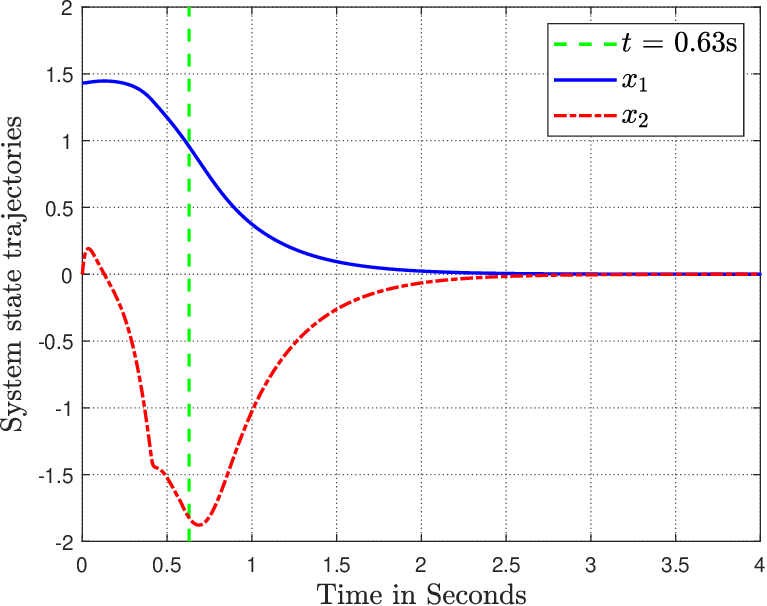}
	\end{minipage}	}
	\subfigure[The control input]{
		\begin{minipage}[t]{0.46\textwidth}
			\centering
			\includegraphics[width=\textwidth]{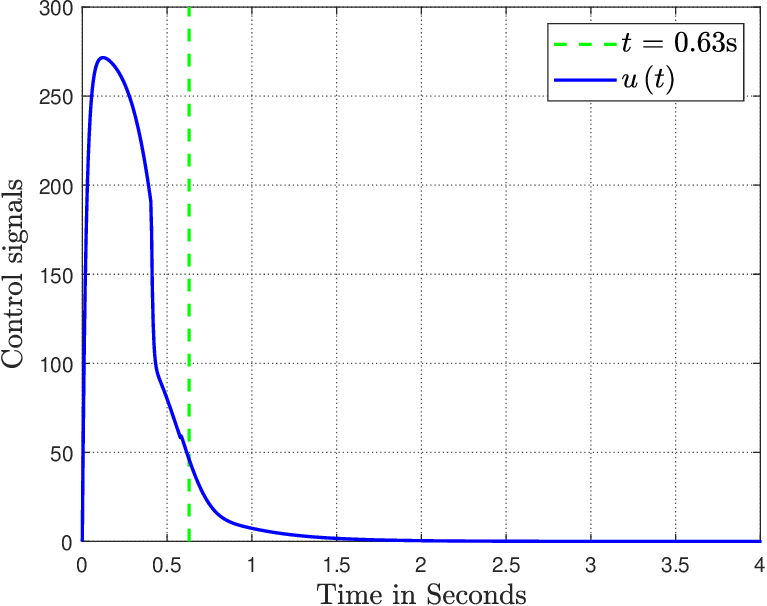}
	\end{minipage}}
	\subfigure[The integral sliding surface variable]{
		\begin{minipage}[t]{0.46\textwidth}
			\centering
			\includegraphics[width=\textwidth]{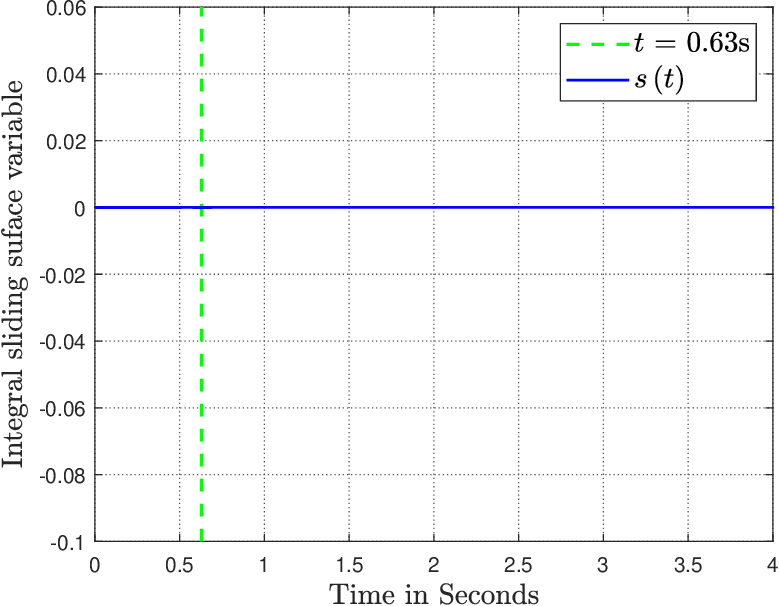}
	\end{minipage}}
	\subfigure[The closed-loop control system trajectories]{
		\begin{minipage}[t]{0.46\textwidth}
			\centering
			\includegraphics[width=\textwidth]{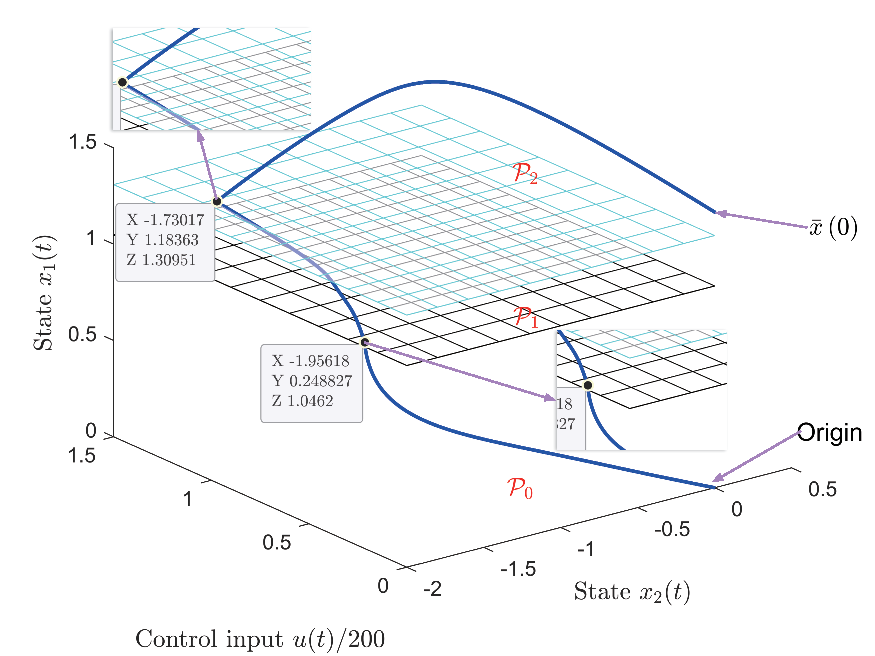}
		\end{minipage}	
	}
	\caption{The numerical experiment}
		\label{ex:invet}
\end{figure}

Similarly, the approximation error bounds are difficult to be obtained precisely. Thus, in this experiment, based on the method in~\cite{gao2013universal}, the norm-bounds of uncertainties are calculated as 
\begin{equation}
\begin{aligned}
\varepsilon _{f_0}&=\max \left\{ \frac{\left\|A_0x\left( t \right)- f(x\left( t \right) ) \right\|}{\left\| x\left( t \right) \right\|},\left\| g(x\left( t \right) )-B_0 \right\| \right\},
	\varepsilon _f=\max \left\{ \frac{1}{2}\frac{\left\| A_ix\left( t \right) +C_i-f(x\left( t \right) ) \right\|}{\left\| x\left( t \right) \right\|},\left\| g(x\left( t \right) )-B_i \right\| \right\}, \\
	\varepsilon _g&=\frac{1}{2}\max \left\| A_ix\left( t \right) +C_i-f(x\left( t \right) ) \right\|, i \in\left\{1,2,3,4\right\}
	\end{aligned}
		\label{re}
\end{equation}
at a series of vertex points 
within the operating region $\left(x_{1}, x_{2}\right) \in\left[ -\frac{\pi}{2},\frac{\pi}{2} \right] \times \left[ -3,3 \right] $. One can calculate that $\varepsilon _{f_0}=0.02$, $\varepsilon _{f}=0.01$, and $\varepsilon _{g}=0.35$ through the numerical calculation of (\ref{re}) on these chosen vertex points.

Base on Algorithm~2, one can obtain the  controller matrices and the sliding surface matrix as follows:
\begin{align*}
	\begin{bmatrix}
		\begin{array}{c|c}
			D_1  &  D_3  \\  \hline
			D_2  &  D_4 
		\end{array}
	\end{bmatrix}
	&=
	\begin{bmatrix}
		\begin{array}{c|c}
			3.00 &  -3.00  \\  \hline 
			5.00 & -5.00  \\  
		\end{array}
	\end{bmatrix}\\
	\bar{K}_0&=\left[ \begin{matrix}
		46381.5662&		13843.0990&		-437.2131\\
	\end{matrix} \right] 
	\\
	\bar{K}_1=\bar{K}_3&=\left[ \begin{matrix}
		13997.0179&		4213.0535&		-133.2537\\
	\end{matrix} \right]
	\\
	\bar{K}_2=\bar{K}_4&=\left[ \begin{matrix}
		8287.5168&		 1620.6117&		-51.5002\\
	\end{matrix} \right]
	\\
	\bar{S}&=\left[ \begin{matrix}
		-0.1269&		-0.0501&		0.00066\\
	\end{matrix} \right] .
\end{align*}

\emph{It is noted that the constructed PALM has different input matrices in each partition. Therefore, the method in~\cite{rubagotti2011integral,li2013adaptive,jiang2018novel} cannot be extended to our case trivially.}

Similarly, the function $\mbox{sgn}\left( s\left( t \right) \right) $ is approximated by the function $$\frac{s\left( t \right)}{\lVert s\left( t \right) \rVert +0.020}.$$

In the simulation, we set  $\bar{x}\left( 0 \right)$ as $\left[ \begin{matrix}82\degree&		0&		0\end{matrix} \right] $. Fig.~\ref{ex:invet}(a) and  Fig.~\ref{ex:invet}(b) show the inverted pendulum trajectories. One can observe that both $x\left(t\right)$ and $u\left(t\right)$ are moving
within the chosen compact region and behave the asymptotic stable sliding motion. In view of highly nonlinearities w.r.t. the inverted pendulum, a large control input is required to restrict system dynamics on the proposed integral sliding manifold and stabilize nonlinear system behaviors.

It is also observed from Fig.~\ref{ex:invet}(c) that the sliding mode is achieved prior to $t=0.63\rm s$ while the system trajectories converge to the origin after $t=3\rm s$. That is, the controlled inverted pendulum enters the sliding mode before the system trajectories are stabilized  to the origin and behaves the ideal system dynamics afterwards. 

It can be observed from the numerical results and Fig.~\ref{ex:invet}(d) that the inverted pendulum trajectories converge to the equilibrium with satisfied performance and no obvious chattering phenomenon appears. Note that the considered system plant has no constant input matrices. The successful application shown in the numerical results demonstrates the advantages of the developed ISMPC scheme.

\begin{figure}[htb]
	\centering
	\subfigure[The inverted pendulum trajectories  move out $X\times U$]{
		\begin{minipage}[t]{0.48\textwidth}
			\centering
			\includegraphics[width=\textwidth]{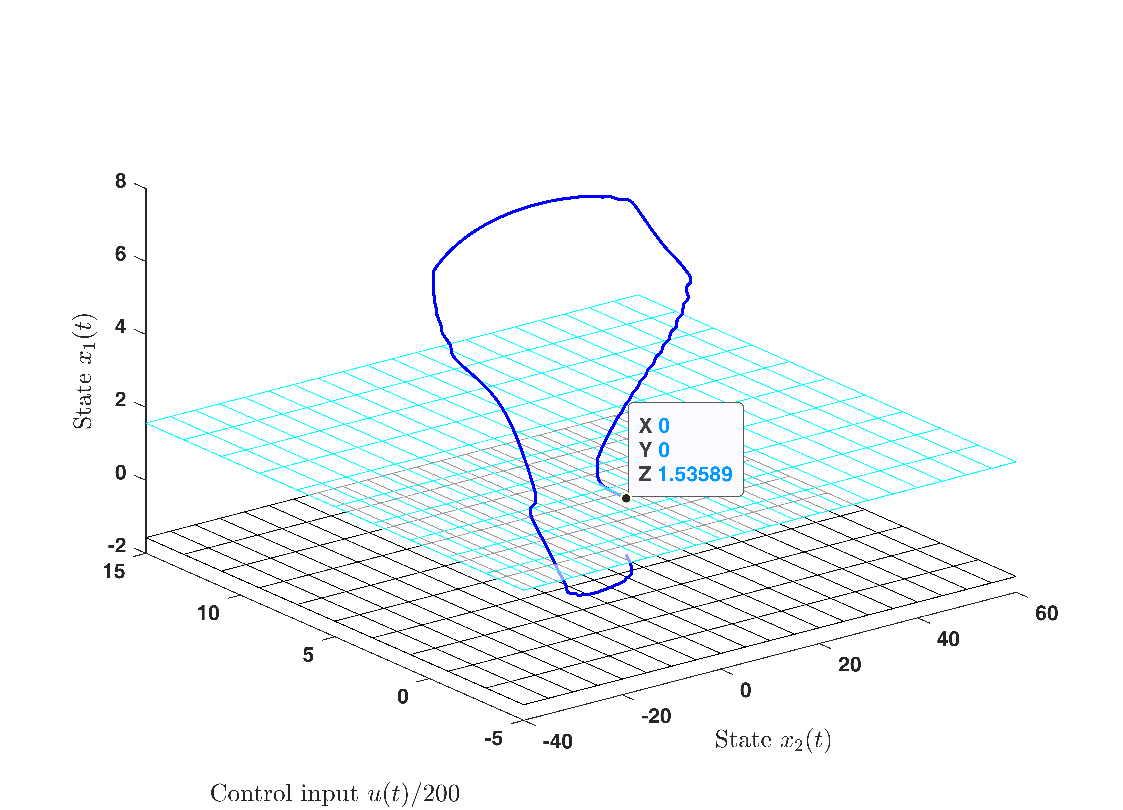}
		\end{minipage}	
	}
	\subfigure[The integral sliding surface is not maintained]{
		\begin{minipage}[t]{0.48\textwidth}
			\centering
			\includegraphics[width=\textwidth]{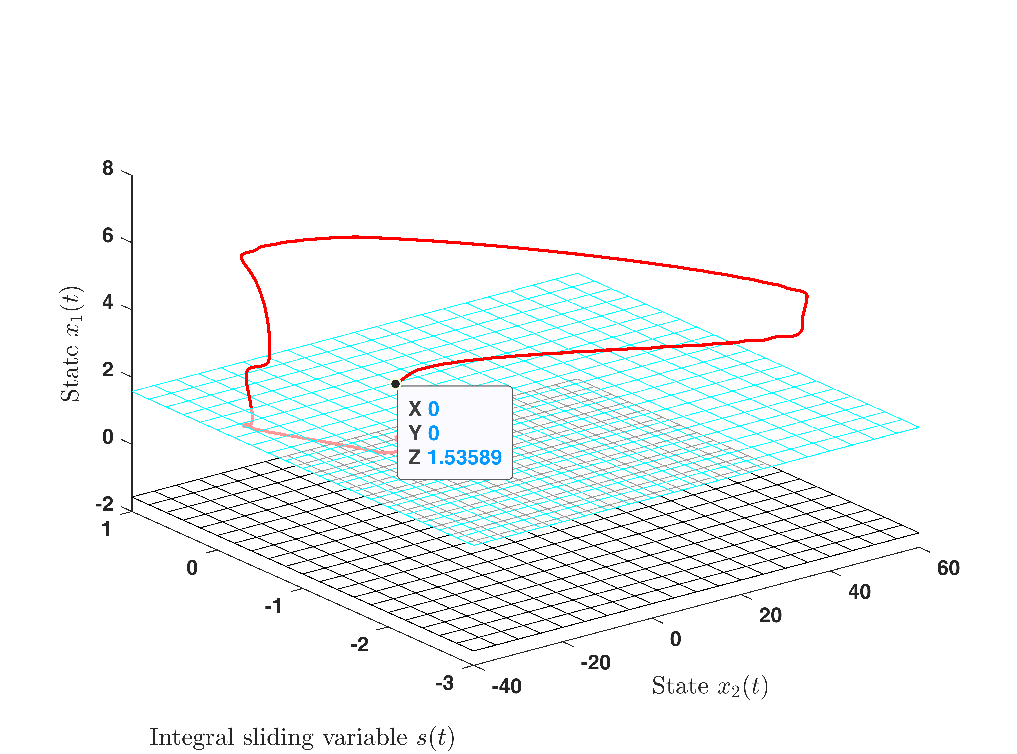}
		\end{minipage}	
	}
	\caption{An undefined case}
	\label{ex:illustrate}
\end{figure}

In order to demonstrate the statement we made in Remark \ref{remark3.1} that the sliding mode can be only achieved ``in potential'', we further consider the case that the system trajectories are initially placed at $\left[ \begin{matrix}88\degree&		0&		0\end{matrix} \right] $, which is very close to the boundary of the region of interest, and the control design results are shown in Fig. \ref{ex:illustrate}. One could observe that the system trajectories move out $X\times U$ and  the ideal stable sliding motion defined in this paper cannot be realized. This is because outside $X\times U$, the approximation PALM in (\ref{pls}) and the piecewise integral sliding surface in (\ref{surf}) are both undefined.

One can also observe from the simulation results of the two examples that first, the developed stabilization strategy is implementable to both nonlinear systems in~(\ref{nonl}) with different input matrices or identical input channel; and second, the proposed ISMPC generates smooth control signals even around the boundary between different subspaces, while the closed-loop control system behaves stable sliding motion.

\section{Conclusions}
\label{sec:conclusion}
A new ISMPC scheme has been proposed to stabilize the general continuous-time non-affine nonlinear systems through PALMs. The proposed control strategy removes a restrictive assumption that is required in relevant research and shows significant convenience in coping with general nonlinear systems by constructing a piecewise integral sliding surface and a corresponding piecewise sliding-mode parallel controller. Moreover, results on the universality of the proposed ISMPC scheme have been provided, which further demonstrates its usefulness. Future research topics include conservatism reduction, universality analysis of different types of controllers and practical applications of the developed method.



%
%
%
\subsection*{FUNDING INFORMATION}
This research was supported by the National Key Research and Development Program of China, Grant/Award Number: 2022YFF0902800; the Natural Science Foundation of China, Grant/Award Number: 61903016, 62273016, U21B6001;  the Alexander von Humboldt Foundation of Germany.

\subsection*{Conflict of interest}
The authors declare no potential conflict of interests.
%
%
%
%

\appendix
\section{A useful Lemma}
\label{sec:a}
\begin{lemma}[S-procedure\cite{boyd1994linear}]
Given some symmetric matrices $S_0,...,S_q\in \mathbb{R}^{n\times n}$, the following conditions on $S_0,...,S_q$,  $\xi ^TS_0\xi >0$, for $\forall  \xi \ne 0$.

$\xi ^TS_i\xi \geqslant 0$, for $i=1,...,q$, are fulfilled when there exists a series of scalars $\tau _1\geqslant 0,...,\tau _q\geqslant 0$, such that $$S_0-\sum_{i=1}^q{\tau _i}S_i>0.$$
	\label{lemmaapen}
\end{lemma}

\section{Proof of Theorem 3}\label{proof:GES}
For a GES nonlinear system in (\ref{nonl}),  a globally exponentially stable system as in~(\ref{clcs}) or~(\ref{nonre}) can be constructed.
In view of the universal approximation capability of the PALM~\cite{julian1998canonical}, for any positive scalars $\varepsilon _f$, $\varepsilon _g$, $\varepsilon _h$ and $\varepsilon _i$, one can  build  the following approximation system w.r.t.~(\ref{clcs}) on $X \times U$:
\begin{equation}
	\left\{ \begin{array}{c}
		\dot{x}\left(t\right)=\hat{f}\left( x,u \right) =f\left( x,u \right) +\varepsilon _f\left( x,u \right) +\varepsilon _g\left( x,u \right)\\
		\dot{u}\left(t\right)=\hat{g}\left( x,u \right) =g\left( x,u \right) +\varepsilon _h\left( x,u \right) +\varepsilon _i\left( x,u \right)\\
	\end{array} \right.
	\label{exnom}
\end{equation}
such that
\begin{align}
	\left\{ \begin{array}{c}
		\lVert \varepsilon _f\left( x,u \right) \rVert \leqslant \varepsilon _f\lVert \left[ x^T,u^T \right] ^T \rVert\\
		\lVert \varepsilon _h\left( x,u \right) \rVert \leqslant \varepsilon _h\lVert \left[ x^T,u^T \right] ^T \rVert\\
		\lVert \varepsilon _g\left( x,u \right) \rVert \leqslant \varepsilon _g\\
		\lVert \varepsilon _i\left( x,u \right) \rVert \leqslant \varepsilon _i\\
	\end{array} \right.
	\label{exerr}
\end{align}
where $f\left( x,u \right)$ and $g\left( x,u \right)$ are defined in~(\ref{clcs}).

In particular, for the partition $\bar X_0$, we have $\max \left\{ \varepsilon _f,\varepsilon _h \right\}\equiv 0$.

The system in~(\ref{exnom}) is rewritten as
\begin{equation}
	\dot{\bar{x}}\left( t \right) =\hat{F}\left( \bar{x}\left( t \right) \right) =F\left( \bar{x}\left( t \right) \right) +\bar{\varepsilon}_f\left( \bar{x}\left( t \right) \right) +\bar{\varepsilon}_g\left( \bar{x}\left( t \right) \right),
	\label{exrenom}
\end{equation}
where
\begin{equation}
	\left\{ \begin{array}{c}
		\hat{F}\left( \bar{x}\left( t \right) \right) =\left[ \hat{f}\left( x,u \right) ^T,\hat{g}\left( x,u \right) ^T \right] ^T\\
		\bar{\varepsilon}_f\left( \bar{x}\left( t \right) \right) =\left[ \varepsilon _f\left(x,u \right) ^T,\varepsilon _h\left(x,u\right) ^T \right] ^T\\
		\bar{\varepsilon}_g\left( \bar{x}\left( t \right) \right) =\left[ \varepsilon _g\left(x,u\right) ^T,\varepsilon _i\left(x,u\right) ^T \right] ^T\\
	\end{array} \right..
	\label{exlmi1}
\end{equation}

Then, the piecewise integral sliding surface is designed as
\begin{equation}
	\label{exsurf}
	s(t)=\bar{S}\left\{\bar{x}(t)-\bar{x}(0)-\int_{0}^{t} \hat{F}(\bar{x}(\theta )) d\theta \right\}=0,
\end{equation}
while the piecewise sliding-mode parallel controller is constructed as
\begin{gather}
	\emph{for}\,\, \bar{x}\left( t \right) \subseteq \bar X_i, i\in \varphi \notag \\
	\dot{u}\left( t \right) =\hat{g}\left(\bar x\left( t \right) \right) -\left( \gamma +\alpha _i +\upsilon _i \left( t \right) \right) S_{u}^{-1}\mbox{sgn}\left( s\left( t \right) \right). 
	\label{expcont}
\end{gather}

It can be concluded from the procedure of Theorem~\ref{thm1} that
(\ref{exsurf}) is reached and maintained from the beginning of evolution and the resultant
sliding motion is
	\begin{equation}
	\dot{\bar{x}}\left( t \right) =\hat{F}\left( \bar{x}\left( t \right) \right) -\left( R_1-R_2S_{u}^{-1}S_x \right) \varepsilon _f\left( x\left( t \right) ,u\left( t \right) \right) -\left( R_1-R_2S_{u}^{-1}S_x \right) \varepsilon _g\left( x\left( t \right) ,u\left( t \right) \right) .
	\label{exsld}
\end{equation}

It follows from the Lyapunov converse theorem in~\cite{khalil2002nonlinear} that, along the trajectories of (\ref{nonre}), there exist a Lyapunov function $V\left( \bar{x}\right) $ and four positive scalars $b_1$, $b_2$, $b_3$, and $b_4$ satisfying
\begin{align}
	\label{clyp1}
	b_1\lVert \bar{x}\rVert ^2\leqslant V\left( \bar{x} \right) &\leqslant b_2\lVert \bar{x} \rVert ^2
	\\
	\label{clyp2}
	\frac{\partial V\left( \bar{x} \right)}{\partial \bar{x}}F\left( \bar{x} \right) &\leqslant -b_3\lVert \bar{x} \rVert ^2
	\\
	\label{clyp3}
	\lVert \frac{\partial V\left( \bar{x}\right)}{\partial \bar{x}} \rVert &\leqslant b_4\lVert \bar{x} \rVert
\end{align}
where $F\left( \bar{x} \right)$ is defined in~(\ref{nonre}).

On the other hand, the derivative of $V\left( \bar{x} \right) $ along the trajectories of (\ref{exrenom}) satisfies
\begin{align}
	\dot{V}\left( \bar{x}\left( t \right) \right) =&\frac{\partial V\left( \bar{x}\left( t \right) \right)}{\partial \bar{x}}\hat{F}\left( \bar{x}\left( t \right) \right) \notag
	\\
	=&\frac{\partial V\left( \bar{x}\left( t \right) \right)}{\partial \bar{x}}F\left( \bar{x}\left( t \right) \right) +\frac{\partial V\left( \bar{x}\left( t \right) \right)}{\partial \bar{x}}\left( \bar{\varepsilon}_f\left( \bar{x}\left( t \right) \right) +\bar{\varepsilon}_g\left( \bar{x}\left( t \right) \right) \right) \notag
	\\
	\leqslant& -b_3\lVert \bar{x}\left( t \right) \rVert ^2+b_4\max \left\{ \varepsilon _f,\varepsilon _h \right\} \lVert \bar{x}\left( t \right) \rVert ^2 +b_4\max \left\{ \varepsilon _g,\varepsilon _i \right\} \lVert \bar{x}\left( t \right) \rVert .
	\label{exlyp}
\end{align}

\emph{Case i:} For the trajectories that move within the partition $\bar X_0$, one has that $\max \left\{ \varepsilon _g,\varepsilon _i \right\} \equiv0$. It then follows from~(\ref{exlyp}) that 
\begin{equation}
	\dot V\left( \bar{x}\left( t \right) \right)\leqslant -(b_3-b_4\max \left\{ \varepsilon _f,\varepsilon _h \right\}) \lVert \bar{x}\left( t \right) \rVert^2
	\label{x1x}
\end{equation}
by choosing the appropriate approximation error bounds $\varepsilon_f$ and $\varepsilon_h$ such that
\begin{align}
	\label{originex}
	\max \left\{ \varepsilon _f,\varepsilon _h \right\}<\frac{b_3}{b_4} .
\end{align}

\emph{Case ii:} For trajectories that move within the partitions $\bar X_i, i \in \varPhi$, one can conclude that the norm $\|\bar x(t)\|$ is lower bounded.

Given a scalar $0<\lambda<1/2 $, one can always choose the approximation error bounds $\varepsilon _g,\varepsilon _i,\varepsilon_f$, and $\varepsilon_h$ such that
\begin{equation}
	\label{lmie1}
	\max \left\{ \varepsilon _g,\varepsilon _i \right\}\leqslant \lambda \frac{b_3}{b_4} \min \lVert \bar{x}\left( t \right) \rVert\leqslant \lambda \frac{b_3}{b_4} \lVert \bar{x}\left( t \right) \rVert
\end{equation}
and
\begin{equation}
	\max \left\{ \varepsilon _f,\varepsilon _h \right\}<(1-\lambda)\frac{b_3}{b_4},
	\label{lmie2}
\end{equation}
which means 
$b_3-b_4\max \left\{ \varepsilon _f,\varepsilon _h \right\}-\lambda b_3>0$.

By submitting~(\ref{lmie1}) and~(\ref{lmie2}) into~(\ref{exlyp}), we have
\begin{equation}
	\dot{V}\left( \bar{x}\left( t \right) \right) \leqslant -b_3\lVert \bar{x}\left( t \right) \rVert ^2+\left(b_4 \max \left\{ \varepsilon _f,\varepsilon _g \right\} +\lambda b_3  \right) \lVert \bar{x}\left( t \right) \rVert ^2. 
	\label{exlyp1}
\end{equation}

From~(\ref{lmie1})-(\ref{exlyp1}),  one has that along the trajectories of~(\ref{exnom}) moving within $\bar X_i, i \in \varPhi$,
\begin{equation}
	\dot{V}\left( \bar{x}\left( t \right) \right) \leqslant -\tilde{b}\lVert \bar{x}\left( t \right) \rVert ^2
	\label{exrelyp}
\end{equation}
where
\begin{equation}
	\tilde{b}=b_3-\left( b_4\max \left\{ \varepsilon _f,\varepsilon _h \right\} +c_3\lambda  \right)>0 .
	\label{exree}
\end{equation}

Therefore,  the semi-global exponential stability of (\ref{exrenom}) is concluded based on
the Lyapunov stability theory and the results in~(\ref{x1x}) and (\ref{exrelyp}).

We are now ready to analyze the stability of (\ref{exsld}) that can be treated as a perturbation system of (\ref{exrenom}). By following the similar procedure of~(\ref{clyp1})–(\ref{exree}) and based on the subsequent facts:
	\begin{align}
	\label{matrixstb1}
	\lVert \left( R_1-R_2S_{u}^{-1}S_x \right) \varepsilon _f\left( x\left( t \right) ,u\left( t \right) \right) \rVert
	\leqslant& \left( 1+\lVert R_2S_{u}^{-1} \rVert \lVert S_x \rVert \right) \varepsilon _f\lVert \bar{x}\left( t \right) \rVert ,\\
	\label{matrixstb2}
	\lVert \left( R_1-R_2S_{u}^{-1}S_x \right) \varepsilon _g\left( x\left( t \right) ,u\left( t \right) \right) \rVert 
	\leqslant& \left( 1+\lVert R_2S_{u}^{-1} \rVert \lVert S_x \rVert \right) \varepsilon _g, \\
	\label{matrix}
	\max\left\{ \varepsilon_g\right\}< \lambda \frac{b_3}{b_4} \min \lVert \bar{x}\left( t \right) \rVert\leqslant &\lambda \frac{b_3}{b_4} \lVert \bar{x}\left( t \right) \rVert,
\end{align}
the semi-global exponential stability of~(\ref{exsld}) is guaranteed, by appropriately designing the approximation error bounds $\varepsilon_f,\varepsilon_h,\varepsilon_g$, and $\varepsilon _i$ such that
\begin{equation}
	\left( 1+\lVert R_2S_{u}^{-1} \rVert \lVert S_x \rVert \right) \varepsilon _f+\max \left\{ \varepsilon _f,\varepsilon _g \right\} <\left[ 1-\left( 2+\lVert R_2S_{u}^{-1} \rVert \lVert S_x \rVert \right) \lambda \right] \frac{b_3}{b_4},
	\label{exstaslm}
\end{equation}
where $$0<\lambda<\frac{1}{2+\lVert R_2S_{u}^{-1} \rVert \lVert S_x \rVert},$$ $b_3$ and $b_4$ are defined in~(\ref{clyp2}) and~(\ref{clyp3}) respectively.
\hfill $\blacksquare$\par

\section{Proof of Theorem 4}\label{proof:GAS}

\indent By using the piecewise sliding-mode parallel controller
in~(\ref{expcont}), we have shown in Appendix.\ref{proof:GES} that the piecewise integral
sliding surface in~(\ref{exsurf}) can be reached and maintained from the beginning of evolution.
So now our main goal is to show that, for a GAS nonlinear
system in~(\ref{nonl}), its corresponding sliding motion in~(\ref{exsld}) will be controlled to
behave the  semi-global asymptotic stability. This can be done with
the aid of $V\left( \bar{x}\left( t \right) \right)$ defined in~(\ref{slyp1})-(\ref{slyp3}).

The derivative of $V\left( \bar{x}\left( t \right) \right)$ defined in~(\ref{slyp1})-(\ref{slyp3}) along the trajectories of~(\ref{exrenom}) satisfies
\begin{align}
	\dot{V}\left( \bar{x}\left( t \right) \right) &=\frac{\partial V\left( \bar{x}\left( t \right) \right)}{\partial \bar{x}}\hat{F}\left( \bar{x}\left( t \right) \right) \notag
	\\
	&=\frac{\partial V\left( \bar{x}\left( t \right) \right)}{\partial \bar{x}}F\left( \bar{x}\left( t \right) \right) +\frac{\partial V\left( \bar{x}\left( t \right) \right)}{\partial \bar{x}}\left( \bar{\varepsilon}_f\left( \bar{x}\left( t \right) \right) +\bar{\varepsilon}_h\left( \bar{x}\left( t \right) \right) \right) \notag
	\\
	&\leqslant -\alpha _3\left( \lVert \bar{x}\left( t \right) \rVert \right) +h\max \left\{ \varepsilon _f,\varepsilon _h \right\} \lVert \bar{x}\left( t \right) \rVert +h\max \left\{ \varepsilon _g,\varepsilon _i \right\}.
	\label{sylyp}
\end{align}

\emph{Case i:} For the trajectories that move within the partition $\bar X_0$, one has
\begin{equation}
	\dot V\left( \bar{x}\left( t \right) \right)\leqslant-\alpha_3(\lVert \bar{x}\left( t \right) \rVert)+\rho\|\bar x(t)\|\leqslant -\alpha_4(\lVert \bar{x}\left( t \right) \rVert),
	\label{x2x}
\end{equation}
if~(\ref{thm5.1}) holds and the norm-bounds of the approximation error $\varepsilon_f$ and $\varepsilon_g$ satisfy
\begin{align}
	\label{originsy}
	\max \left\{ \varepsilon _f,\varepsilon _h \right\}<\frac{\rho}{h}.
\end{align}

\emph{Case ii:} For trajectories that move within the partitions $\bar X_i, i \in \varPhi$, the norm $\|\bar x(t)\|$ is lower bounded.

Given a scalar $0<\mu <1/2$, one can always choose the approximation error bounds $\varepsilon _g,\varepsilon _i, \varepsilon_f$, and $\varepsilon_h$ such that
\begin{equation}
	\label{sylmie1}
	\max \left\{ \varepsilon _g,\varepsilon _i \right\} \leqslant \mu \frac{\rho}{h} \min \lVert \bar{x}\left( t \right) \rVert
	\leqslant \mu \frac{\rho}{h} \|\bar x(t)\|
\end{equation}
and
\begin{align}
	\max \left\{ \varepsilon _f,\varepsilon _h \right\}<(1-\mu)\frac{\rho}{h},
	\label{sylmie2}
\end{align}
which means 
\begin{equation}
	h\max \left\{ \varepsilon _f,\varepsilon _h \right\}+\mu\rho<\rho .
	\label{x3x}
\end{equation}

By submitting~(\ref{sylmie1})-~(\ref{x3x}) into~(\ref{sylyp}), one has
\begin{align}
	\dot{V}\left( \bar{x}\left( t \right) \right) &\leqslant -\alpha _3\left( \lVert \bar{x}\left( t \right) \rVert \right) +\left( h\max \left\{ \varepsilon _f,\varepsilon _h \right\} +\mu\rho \right) \lVert \bar{x}\left( t \right) \rVert \notag \\
	&\leqslant -\alpha_3(\lVert \bar{x}\left( t \right) \rVert)+\rho\|\bar x(t)\|\notag \\ 
	&\leqslant -\alpha_4(\lVert \bar{x}\left( t \right) \rVert),
	\label{sylyp1}
\end{align}
if~(\ref{thm5.1}) holds.
The semi-global asymptotic stability of~(\ref{exnom}) is then concluded by following from both~(\ref{x2x}) and~(\ref{sylyp1}).

The sliding motion in~(\ref{exsld}) can be treated as a perturbation system of~(\ref{exrenom}). Similar with the procedure of~(\ref{sylyp})-(\ref{sylyp1}), by choosing the norm-bounds of the approximation error  $\varepsilon_f,\varepsilon_h,\varepsilon_g$, and $\varepsilon _i$ such that
\begin{equation}
	\left( 1+\lVert R_2S_{u}^{-1} \rVert \lVert S_x \rVert \right) \varepsilon _f+\max \left\{ \varepsilon _f,\varepsilon _g \right\} 
	<\left[ 1-\left( 2+\lVert R_2S_{u}^{-1} \rVert \lVert S_x \rVert \right) \mu \right] \frac{\rho}{h},
	\label{systaslm}
\end{equation}
where $$0<\mu<\frac{1}{2+\lVert R_2S_{u}^{-1} \rVert \lVert S_x \rVert},$$
$\rho$ and $h$ are defined in~(\ref{slyp3}) and~(\ref{thm5.1}), respectively, one can conclude that the~(\ref{exsld}) is semi-globally asymptotically stable. \hfill $\blacksquare$\par 

\bibliography{ISMPC-IJRNC}%

\begin{thebibliography}{10}
\providecommand \doibase [0]{http://dx.doi.org/}%

\bibitem{johansson2003piecewise}
Johansson MKJ. {\it Piecewise Linear Control Systems: A Computational
  Approach}.
\newblock Springer .
\newblock 2003.

\bibitem{kersting2017direct}
Kersting S, Buss M. Direct and indirect model reference adaptive control for
  multivariable piecewise affine systems. {\it IEEE Transactions on Automatic
  Control} 2017\string; 62(11)\string: 5634--5649.

\bibitem{li2018stability}
Li P, Lam J, Kwok KW, Lu R. Stability and stabilization of periodic piecewise
  linear systems: {A} matrix polynomial approach. {\it Automatica} 2018\string;
  94\string: 1--8.

\bibitem{julian1998canonical}
Julian P, Jordan M, Desages A. Canonical piecewise-linear approximation of
  smooth functions. {\it IEEE Transactions on Circuits and Systems I:
  Fundamental Theory and Applications} 1998\string; 45(5)\string: 567-571.

\bibitem{qiu2011approaches}
Qiu J, Feng G, Gao H. Approaches to robust $\mathscr{H}_{\infty}$ static output
  feedback control of discrete-time piecewise-affine systems with norm-bounded
  uncertainties. {\it International Journal of Robust and Nonlinear Control}
  2011\string; 21(7)\string: 790--814.

\bibitem{zhang2016observer}
Zhang L, Ning Z, Zheng WX. Observer-based control for piecewise-affine systems
  with both input and output quantization. {\it IEEE Transactions on Automatic
  Control} 2016\string; 62(11)\string: 5858--5865.

\bibitem{wei2018new}
Wei Y, Yu H, Karimi HR, Joo YH. New approach to fixed-order output-feedback
  control for piecewise-affine systems. {\it IEEE Transactions on Circuits and
  Systems I: Regular Papers} 2018\string; 65(9)\string: 2961--2969.

\bibitem{desimini2020robust}
Desimini R, Prandini M. Robust constrained control of piecewise affine systems
  through set-based reachability computations. {\it International Journal of
  Robust and Nonlinear Control} 2020\string; 30(15)\string: 5989--6020.

\bibitem{xu2022passivity}
Xu N, Zhu Y, Chen X, Su CY. Passivity-based adaptive fault-tolerant control for
  continuous-time Markov jump PWA systems with actuator faults. {\it
  International Journal of Robust and Nonlinear Control} 2022\string;
  32(4)\string: 2300--2312.

\bibitem{utkin1977variable}
Utkin V. Variable structure systems with sliding modes. {\it IEEE Transactions
  on Automatic Control} 1977\string; 22(2)\string: 212--222.

\bibitem{razmi2019neural}
Razmi H, Afshinfar S. Neural network-based adaptive sliding mode control design
  for position and attitude control of a quadrotor {UAV}. {\it Aerospace
  Science and Technology} 2019\string; 91\string: 12--27.

\bibitem{wang2019new}
Wang Y, Feng Y, Zhang X, Liang J. A new reaching law for antidisturbance
  sliding-mode control of {PMSM} speed regulation system. {\it IEEE
  Transactions on Power Electronics} 2019\string; 35(4)\string: 4117--4126.

\bibitem{wang2020usde}
Wang S, Tao L, Chen Q, Na J, Ren X. {USDE}-based sliding mode control for servo
  mechanisms with unknown system dynamics. {\it IEEE/ASME Transactions on
  Mechatronics} 2020\string; 25(2)\string: 1056--1066.

\bibitem{LIU2020108596}
Liu Z, Karimi HR, Yu J. Passivity-based robust sliding mode synthesis for
  uncertain delayed stochastic systems via state observer. {\it Automatica}
  2020\string; 111\string: 108596.

\bibitem{wei2021dynamic}
Wei Y, Karimi HR. Dynamic sliding mode control for nonlinear parameter-varying
  systems. {\it International Journal of Robust and Nonlinear Control}
  2021\string; 31(17)\string: 8408--8419.

\bibitem{zhang2021design}
Zhang J, Shi D, Xia Y. Design of sliding mode output feedback controllers via
  dynamic sliding surface. {\it Automatica} 2021\string; 124\string: 109310.

\bibitem{zhang2021finite}
Zhang J, Wu ZG, Xia Y. Finite-time composite disturbance rejection control for
  discrete-time {Markovian} jump systems. {\it IEEE Transactions on Automatic
  Control} 2022\string; 67(12)\string: 6866-6872.

\bibitem{utkin1997integral}
Utkin V, Shi J. Integral sliding mode in systems operating under uncertainty
  conditions. In: the 35th IEEE Conference on Decision and Control. IEEE. ;
  1996; Kobe, {Japan}\string: 4591-4596.

\bibitem{gao2014suniversal}
Gao Q, Liu L, Feng G, Wang Y. Universal fuzzy integral sliding-mode controllers
  for stochastic nonlinear systems. {\it IEEE Transactions on Cybernetics}
  2014\string; 44(12)\string: 2658--2669.

\bibitem{pan2017integral}
Pan Y, Yang C, Pan L, Yu H. Integral sliding mode control: {Performance},
  modification, and improvement. {\it IEEE Transactions on Industrial
  Informatics} 2017\string; 14(7)\string: 3087--3096.

\bibitem{gao2019fault}
Gao Y, Liu J, Sun G, Liu M, Wu L. Fault deviation estimation and integral
  sliding mode control design for {Lipschitz} nonlinear systems. {\it Systems
  \& Control Letters} 2019\string; 123\string: 8--15.

\bibitem{li2021a}
Li J, Zhai D. A descriptor regular form-based approach to observer-based
  integral sliding mode controller design. {\it International Journal of Robust
  and Nonlinear Control} 2021\string; 31(11)\string: 5134-5148.

\bibitem{ZHANG2022491}
Zhang C, Gong D, Gao Q, Chen W, Wang J. A fuzzy integral sliding-mode parallel
  control approach for nonlinear descriptor systems. {\it Information Sciences}
  2022\string; 615\string: 491-503.

\bibitem{cao1997analysis}
Cao SG, Rees NW, Feng G. Analysis and design for a class of complex control
  systems part {I}: {Fuzzy} modelling and identification. {\it Automatica}
  1997\string; 33(6)\string: 1017--1028.

\bibitem{ho2007robust}
Ho DW, Niu Y. Robust fuzzy design for nonlinear uncertain stochastic systems
  via sliding-mode control. {\it IEEE Transactions on Fuzzy Systems}
  2007\string; 15(3)\string: 350--358.

\bibitem{rubagotti2011integral}
Rubagotti M, Estrada A, Casta{\~n}os F, Ferrara A, Fridman L. Integral sliding
  mode control for nonlinear systems with matched and unmatched perturbations.
  {\it IEEE Transactions on Automatic Control} 2011\string; 56(11)\string:
  2699--2704.

\bibitem{li2013adaptive}
Li H, Yu J, Hilton C, Liu H. Adaptive sliding-mode control for nonlinear active
  suspension vehicle systems using {T–S} fuzzy approach. {\it IEEE
  Transactions on Industrial Electronics} 2013\string; 60(8)\string: 3328-3338.

\bibitem{jiang2018novel}
Jiang B, Karimi HR, Kao Y, Gao C. A novel robust fuzzy integral sliding mode
  control for nonlinear semi-{Markovian} jump {T-S} fuzzy systems. {\it IEEE
  Transactions on Fuzzy Systems} 2018\string; 26(6)\string: 3594--3604.

\bibitem{xi2010piecewise}
Xi Z, Feng G, Hesketh T. Piecewise integral sliding-mode control for {T-S}
  fuzzy systems. {\it IEEE Transactions on Fuzzy Systems} 2010\string;
  19(1)\string: 65--74.

\bibitem{gao2013universal}
Gao Q, Liu L, Feng G, Wang Y, Qiu J. Universal fuzzy integral sliding-mode
  controllers based on {T-S} fuzzy models. {\it IEEE Transactions on Fuzzy
  Systems} 2013\string; 22(2)\string: 350--362.

\bibitem{teixeira1998stabilizing}
Teixeira M, Zak S. Stabilizing controller design for uncertain nonlinear
  systems using fuzzy models. {\it IEEE Transactions on Fuzzy Systems}
  1999\string; 7(2)\string: 133-142.

\bibitem{rodrigues2005piecewise}
Rodrigues L, Boyd S. Piecewise-affine state feedback for piecewise-affine slab
  systems using convex optimization. {\it Systems \& Control Letters}
  2005\string; 54(9)\string: 835--853.

\bibitem{tong2013adaptive}
Tong S, Wang T, Li Y, Zhang H. Adaptive neural network output feedback control
  for stochastic nonlinear systems with unknown dead-zone and unmodeled
  dynamics. {\it IEEE Transactions on Cybernetics} 2013\string; 44(6)\string:
  910--921.

\bibitem{wen2017adaptive}
Wen S, Chen MZ, Zeng Z, Huang T, Li C. Adaptive neural-fuzzy sliding-mode
  fault-tolerant control for uncertain nonlinear systems. {\it IEEE
  Transactions on Systems, Man, and Cybernetics: Systems} 2017\string;
  47(8)\string: 2268--2278.

\bibitem{yang2020event}
Yang D, Li T, Xie X, Zhang H. Event-triggered integral sliding-mode control for
  nonlinear constrained-input systems with disturbances via adaptive dynamic
  programming. {\it IEEE Transactions on Systems, Man, and Cybernetics:
  Systems} 2020\string; 50(11)\string: 4086-4096.

\bibitem{boyd1994linear}
Boyd S, El~Ghaoui L, Feron E, Balakrishnan V. {\it Linear Matrix Inequalities
  in System and Control Theory}.
\newblock Philadelphia, PA, USA: SIAM .
\newblock 1994.

\bibitem{zhang2017event}
Zhang C, Hu J, Qiu J, Chen Q. Event-triggered nonsynchronized $\mathcal
  {H}_{\infty }$ filtering for discrete-time {T–S} fuzzy systems based on
  piecewise {Lyapunov} functions. {\it IEEE Transactions on Systems, Man, and
  Cybernetics: Systems} 2017\string; 47(8)\string: 2330-2341.

\bibitem{wang2020a}
Wang M, Qiu J, Feng G. A novel piecewise affine filtering design for {T–S}
  fuzzy affine systems using past output measurements. {\it IEEE Transactions
  on Cybernetics} 2020\string; 50(4)\string: 1509-1518.

\bibitem{farbood2021fuzzy}
Farbood M, Shasadeghi M, Niknam T, Safarinejadian B. Fuzzy {Lyapunov}-based
  model predictive sliding-mode control of nonlinear systems: {An} ellipsoid
  recursive feasibility approach. {\it IEEE Transactions on Fuzzy Systems}
  2022\string; 30(6)\string: 1929--1938.

\bibitem{wang2014outage}
Wang KY, So AMC, Chang TH, Ma WK, Chi CY. Outage constrained robust transmit
  optimization for multiuser {MISO} downlinks: {Tractable} approximations by
  conic optimization. {\it IEEE Transactions on Signal Processing} 2014\string;
  62(21)\string: 5690--5705.

\bibitem{khalil2002nonlinear}
Khalil HK, Grizzle JW. {\it Nonlinear Systems}.
\newblock Des Moines, IA, USA: Prentice Hall .
\newblock 2002.

\bibitem{chua1986canonical}
Chua L, Deng A. Canonical piecewise-linear modeling. {\it IEEE Transactions on
  Circuits and Systems} 1986\string; 33(5)\string: 511--525.

\bibitem{zhang2007output}
Zhang T, Feng G. Output tracking of piecewise-linear systems via error feedback
  regulator with application to synchronization of nonlinear {Chua's} circuit.
  {\it IEEE Transactions on Circuits and Systems I: Regular Papers}
  2007\string; 54(8)\string: 1852--1863.

\end{thebibliography}

\end{document}